\documentclass[prd,preprint,showpacs,preprintnumbers,amsmath,amssymb]{revtex4}

\usepackage{graphicx}% Include figure files
\usepackage{dcolumn}% Align table columns on decimal point
\usepackage{bm}% bold math

\begin{document}

\preprint{UTHET-09-0201}

\title{Particle emission from a black hole on a tense codimension-2 brane}% Force line breaks with \\

\author{Usama A.~al-Binni}
\email{ualbinni@tennessee.edu}
\author{George Siopsis}%
 \email{siopsis@tennessee.edu}
\affiliation{%
Department of Physics and Astronomy,
The University of Tennessee,
Knoxville, TN 37996 - 1200, USA.
}%

\date{February 2009}% It is always \today, today,

\begin{abstract}
We calculate analytically grey-body factors of
Schwarzschild black-holes localized on a 3-brane of finite tension and codimension 2.
We obtain explicit expressions  for various
types of particles emitted in the bulk as well as on the brane in both the low and high frequency regimes.
In the latter case, we obtain expressions which are valid for arbitrary number of extra dimensions if the brane tension vanishes.
\end{abstract}

\maketitle

\section{Introduction}

The possibility of producing mini black holes in high energy
collisions is one of the exciting prospects of research at the LHC.
It appears as one of the major implications of brane-world models,
where the presence of extra dimensions opens up the possibility for
a lowered Planck scale $M_*$
\cite{bibADD1,bibADD2,bibADD3,bibRS1,bibRS2}, so that gravity is
much stronger in the spacetime which includes the extra dimensions
and black holes can then be produced in high energy collisions
\cite{bibBanks,bibADM,bibDL,bibGT}.

Two key signatures of such black holes would be the observation of Hawking radiation and the measurement of quasi-normal
mode frequencies
\cite{bibKS,bibSM2,bibNoll,bibNatario,bibHarmark}. Although at the
horizon of a black hole the emitted Hawking radiation is a perfect
black-body radiation, for an observer at infinity the radiation
appears modified.
This is because in escaping from the black hole
the radiation needs to cross the non-trivial potential around the
black hole which implies that part of the radiation will
backscatter, resulting in a frequency-dependent factor that modifies
the otherwise ideal black body radiation. This is known as a
grey-body radiation and has a spectrum described by a modified
black-body emission rate, which in $D=4+n$ dimensions is of the form
\begin{equation}\label{eq1}
\frac{{{\text{d}}E(\omega )}} {{{\text{d}}t}} = \sum\limits_{\ell
,\kappa} {\sigma _{\ell ,\kappa} (\omega )\frac{\omega } {{e^{\omega
/T_{\text{H}} } \mp 1}}\frac{{{\text{d}}^{n + 3} k}} {{(2\pi )^{n +
3} }}}
\end{equation}
where $\ell $ stands for the angular momentum quantum number, and
$\kappa$ for any other quantum numbers of the emitted particles.
$\sigma _{\ell ,\kappa }(\omega )$ is the grey-body factor modifying
the black-body radiation at the horizon to the spectrum observed at
asymptotic infinity away from the black hole. The Hawking
temperature $T_{\mathrm{H}}$ for a Schwarzschild black-hole is given
in terms of the horizon radius $r_h$ and the number of extra
dimensions $n$ \cite{bibMyers},
\begin{equation}\label{eq2}
T_{\text{H}}  = \frac{{n + 1}} {{4\pi r_h }}
\end{equation}
For such a black hole, $r_h$ is enhanced and given by
\begin{equation}\label{eq2a}
r_h  = \frac{1} {{\sqrt \pi  M_* }}\left( {\frac{{M_{\mathrm{BH}} }}
{{M_* }}} \right)^{\frac{1} {{n + 1}}} \left( {\frac{{8\Gamma \left(
{\frac{{n + 3}} {2}} \right)}} {{n + 2}}} \right)^{\frac{1} {{n +
1}}}
\end{equation}
where $M_{\mathrm{BH}}$ is the mass of the black-hole.

Clearly, an experimental observation of Hawking radiation from such
black holes is a very exciting prospect as it constitutes direct
observational evidence of black-holes. It should provide a rich
source of information about the dimensionality and structure of
spacetime as well as constitute a first step toward establishing
quantum gravity as a discipline accessible to experimental
investigation. One way of exploring experimental implications of
such theories that can be observed at accelerators has been through
event generators. Many such programs have appeared over the years
\cite{Snowmass,bibHRW,bibCGCS}, the most recent and comprehensive of
which is perhaps {\sf BlackMax} \cite{bibDSSC}. Based on
phenomenologically realistic models free of serious problems that
plague low-scale gravity, it incorporates all grey-body factors
known to date along with several effects that make predictions which are as
realistic as possible.

In brane-world models, we need to localize Standard Model (SM)
particles on a 3-brane yet allow gravity full access to the bulk.
Such a localization is often done via a projection from higher
dimensional spacetime onto a four-dimensional hypersurface
identified as a 3-brane representing our world. While this method
succeeds in describing the geometry, it fails to capture the
defining feature of the brane, which is its tension. The latter is
often neglected owing to the difficulties in obtaining exact
solutions of Einstein's equations in higher dimensions in the
presence of a brane of finite tension. Recently, a solution that
incorporates a 3-brane of finite tension in six-dimensional
spacetime was constructed by Kaloper and Kiley for static black
holes \cite{bibKK}. It was subsequently extended to include rotating
black holes \cite{bibKiley}. In this model, at the position of the
brane a conical singularity forms that acts to off-load tension
thereby keeping both bulk and brane locally flat. Some work
\cite{bibDKSS,bibCWS,bibCWSWP,bibKNT,bibCCDN4} has already been done
to study the implications of including tension in this way,
resulting in qualitative as well as quantitative modifications which
may affect the interpretation of experimental results.

The present work aims at continuing in that line by studying
analytically the effects of including tension on grey-body factors.
In detail, Section \ref{sec2} is a quick review of the main effects
of including tension in the Schwarzschild case and the attendant
modifications of the parameters of the black-hole. In Section
\ref{sec3} we obtain analytical expressions for low-frequency
grey-body factors for bulk as well as brane-localized emissions, and
and then compare with exact numerical results. In Section \ref{sec4}
we calculate analytically grey-body factors in the case of large
imaginary frequencies, again for emissions both in the bulk and on
the brane. In the tensionless limit, our expressions are valid for
arbitrary number of extra dimensions. Finally, Section \ref{sec5}
contains our conclusions.

\section{Schwarzschild black hole on a tense codimension-2 3-brane}
\label{sec2}

Black-holes formed through high energy collisions are expected to
evaporate through several phases (e.g. \cite{bibGT}). After the
initial ``balding'' phase where the gauge hair and other asymmetries
are lost, a spin-down phase ensues, with a small portion of mass
being lost by radiating away the angular momentum. However, it is
expected that the black-hole will spend most of its life in the
Schwarzschild phase emitting spherically symmetric Hawking
radiation. In what follows we will concentrate only on this phase.

In brane-world scenarios, a small black hole with radius $r_h$ which
is much smaller than the size of the extra dimensions $R$ is in fact
completely immersed in the full $4+n$ dimensional spacetime. In
asymptotically flat space, the metric that describes a non-rotating
black-hole in higher dimensions is a generalization of the
4-dimensional Schwarzschild metric \cite{bibMyers},
\begin{equation}\label{eq3}
\mathrm{d} s^2=-f_n(r)\mathrm{d} t^2+\frac{1}{f_n(r)}\mathrm{d} r^2+r^2\mathrm{d}
\Omega _{2+n}^2,\quad f_n(r)=1-\left(\frac{r_h}{r}\right){}^{n+1}
\end{equation}
where the element $\mathrm{d} \Omega _{2+n}$ defines a
$(2+n)$-sphere. To geometrically localize fields on the brane, what
has often been done is to project onto the brane by setting all
angular coordinates $\theta _i$ to $\pi/2$, except for the usual
4-dimensional angles $\theta $ and $\varphi $ (see, e.g.,
\cite{bibKantiI,bibKantiII}). The resulting 4D metric then reads
\begin{equation}\label{eq6}
\mathrm{d} s^2=-f_n(r)\mathrm{d} t^2+\frac{1}{f_n(r)}\mathrm{d}
r^2+r^2\mathrm{d} \Omega _2^2
\end{equation}
As mentioned earlier, such treatments neglecting the tension $\lambda$ of the brane
are mainly due to the the difficulty of obtaining higher-dimensional
solutions of Einstein's equations in the presence of a brane of finite tension.
They can be justified on the grounds that in order for a
black hole to be treated semi-classically, its mass must be
much bigger than the Planck scale, so that the tension of the brane does not appreciably alter
the black hole background and at a scale $\mathcal{O}(r_h)$ the solution is
flat. While this is in principle true, it turns out that for a black
hole to be clear of the quantum regime in this context, given a
gravity scale of the order of $1\operatorname{TeV}$, the mass of the
black hole needs to be on the order of just a few TeV
\cite{bibGT}. However, because brane tension will be generated from
contributions to vacuum fluctuations due to brane-localized
fields, one expects the tension to be of the same order as the
fundamental scale of the theory. Therefore, in order to be
realistic, the brane tension ought to be taken into account.

By introducing tension using the Kaloper-Kiley codimension-2 setup
\cite{bibKK}, the tension of the brane does not curve either the
brane or the bulk, but forms a conical singularity instead, creating
a deficit angle $\Delta \psi $. By adapting to brane tension,
$\Delta \psi $ provides a self-tuning mechanism which leaves both
bulk and brane locally flat. It is given by
\begin{equation}\label{eq7}
\Delta \psi =2\pi (1-b)\lambda  M_*^4
\end{equation}
where $ 0<b\leq 1$. The resulting geometry is the same as that of
Schwarzschild, except that a wedge defining the deficit angle is now
cut out with its edges identified, or equivalently, the coordinate
$\psi$ gets rescaled. The resulting metric reads
\begin{equation}\label{eq8}
\mathrm{d}s_6^2=-f(r)\mathrm{d} t^2+f(r)^{-1}\mathrm{d} r^2+r^2
\mathrm{d}\Omega_4^2,\quad f(r)=1-\left(\frac{r_h}{r}\right){}^3
\end{equation}
with
\begin{equation}\label{eq4}
\mathrm{d}\Omega_4^2=\mathrm{d}\theta^{2}+\sin^{2}\theta \left[
\mathrm{d}\varphi^2+\sin^{2}\varphi (\mathrm{d}\chi ^{2}+
b^{2}\sin^{2}\chi \mathrm{d}\psi ^{2}) \right]
\end{equation}
One significant implication that can be readily shown is that the
radius of the horizon gets enhanced,
\begin{equation}\label{eq10}
r_h=\left(\frac{\mu }{b}\right)^{1/3},\qquad \mu
=\frac{M_{\text{BH}}}{4\pi ^2M_*^4}
\end{equation}
hence also the geometrical cross-section and production
rate, which is important from an
experimental point of view.

Semi-classically, we can study the propagation of particles of spin
$s$ in the vicinity of a black hole by writing the corresponding
field equation in the background of the metric of the black hole.
The resulting equation can then be separated into a radial equation
and a set of angular equations. The radial equation is most
conveniently dealt with by casting it into a Schr\"odinger-like
form,
\begin{equation}\label{eq11}
\frac{\mathrm{d}^2\Psi \left(r_*\right)}{\mathrm{d}
r_*^2}+\left(\omega ^2-V\left[r\left(r_*\right)\right]\right)\Psi
\left(r_*\right)=0
\end{equation}
where $r_*$ is the ``tortoise coordinate'' defined by
\begin{equation}\label{eq12}
r_*=\int \frac{\mathrm{d}r}{f(r)}
\end{equation}
and $V(r)$ is an effective potential which depends on the type of
particle considered. The potential also depends on the eigenvalues
of the angular equations, so that in order to find the
observationally significant eigenvalue $\omega $ of the radial
equation, we need to first solve the angular eigenvalue problems. An
exact solution for the angular eigenvalues for integer spin was
first obtained in \cite{bibPaper1}. It was subsequently derived
using Jacobi polynomials for Dirac fermions in \cite{bibCCDN1}.

We can sum up the effects of including tension in the codimension-2
model for emissions in the bulk by the following replacement rules
\begin{equation}\label{eq5}
r_h\to \frac{r_h}{\sqrt[3]{b}},\quad \ell _i\to \lambda _i=\ell
_i+\left(\frac{1}{b}-1\right)m,\quad \psi \to b \psi
\end{equation}
where $\ell_i$ is the orbital quantum number for the $i$th angular
equation, and $m$ is the magnetic quantum number. For
brane-localized emissions the same rules apply except that the
orbital quantum number does not get modified. In subsequent sections
these rules will be the main ingredient in obtaining explicit
analytical expressions for grey-body factors including brane
tension.

\section{Low-frequency grey-body factors}
\label{sec3}

Calculations of grey-body in the case of low-frequency have been
performed both for bulk
\cite{bibKantiI,bibCEKT,bibCCG1} and brane-localized emissions
\cite{bibKantiII,bibIda1,bibIda2}. In this section we shall extend these results to include brane tension.

In general, one tries to find an approximate solution
of the radial equation which is valid at infinity and another one valid near the horizon.
Then by matching the two solutions in the common domain of validity one can extract the coefficients
of reflection and transmission and hence deduce the absorption
coefficient (grey-body factor) as a function of frequency.

For bulk emissions of integer spin, we can use the Ishibashi-Kodama
`master equation' \cite{bibIK,bibCCG1}, in which the potential in
$D=n+4$ dimensions is given by
\begin{equation}\label{eq9}
V=f(r)\left[\frac{\ell (\ell
+D-3)}{r^2}+\frac{(D-2)(D-4)}{4r^2}+\frac{\left(1-p^2\right)(D-2)^2}{4}\frac{r_h^{D-3}}{r^{D-1}}\right]
\end{equation}
with $p$ determining the type of field (particle) one is interested in
\begin{equation}\label{eq13}
p=\left\{
\begin{array}{ccl}
 0 & , & \text{scalar and gravitational tensor} \\
 2 & , & \text{gravitational vector} \\
 2/(D-2) & , & \text{EM (gauge) vector} \\
 2(D-3)/(D-2) & , & \text{EM (scalar)}
\end{array}
\right.
\end{equation}
and $\ell$ being the angular momentum quantum number taking on values
\begin{equation}\label{eq14}
\ell =\left\{
\begin{array}{ccl}
 0,1,\dots & , & \text{scalar} \\
 1,2,\dots & , & \text{EM} \\
 2,3,\dots & , & \text{gravitational}
\end{array}
\right.
\end{equation}
Gravitational scalar fields are governed by a more complicated potential,
but they can be described approximately using the above potential by
taking $p$ to be \cite{bibCCG1}
\begin{equation}\label{eq15}
p_{\text{grav-scalar}}\sim 2+0.674D^{-0.5445}
\end{equation}
Specifically, for the case we are interested in, $n=2$ we have
\begin{equation} p_{\text{grav-scalar}}=2.25
\end{equation}
Writing the radial equation in terms of $R(r)=r^{-1-\frac{n}{2}}\Psi \left(r_*\right)$ for $n=2$, we end up
with
\begin{equation}\label{eq16}
\frac{f}{r^4}\frac{\mathrm{d}}{\mathrm{d} r}\left[r^4f
R'(r)\right]+\left[\omega ^2-f\left(\frac{\lambda _3\left(3+\lambda
_3\right)}{r^2}-4p^2\frac{1}{r^5}r_h^3\right)\right]R(r)=0
\end{equation}
where we have replaced $\ell$ by $\lambda
_3$ (eq.~(\ref{eq5})) in order to account for the brane tension \cite{bibPaper1}.
Near the horizon, it is
convenient to make the change of variables $r\to f(r)$ (note that $f(r)\to 0$
as $r\to r_h$). The resulting equation is
\begin{equation}\label{eq17}
(1-f)\left(f \frac{\mathrm{d}^2}{\mathrm{d}
f^2}R_{NH}+\frac{\mathrm{d}}{\mathrm{d}
f}R_{NH}\right)+\left(\frac{\left(\omega
r_h\right){}^2}{9f(1-f)}-\frac{\lambda _3\left(3+\lambda
_3\right)}{9(1-f)}+\frac{4}{9}p^2\right)R_{NH}=0
\end{equation}
whose general solution can be written in terms of hypergeometric functions,
\begin{equation}\label{eq18}
R_{\text{NH}}(f)=A_-f^{\alpha}(1-f){}^{\beta
}F(a,b,c;f)+A_+f^{-\alpha}(1-f){}^{\beta
}F(a-c+1,b-c+1,2-c;f)
\end{equation}
where
\begin{equation}\label{eq19}
\begin{aligned}
\alpha & =\frac{i \omega  r_h}{3},\quad \beta =\frac{1}{2}- \frac{1}{3}\sqrt{\left(\lambda
_3+\frac{3}{2}\right)^2-\left(\omega  r_h\right)^2},\\
a & =\alpha +\beta -\frac{2}{3}p=-\frac{i \omega r_h
}{3}+\frac{1}{2}-\frac{1}{3}\sqrt{\left(\lambda
_3+\frac{3}{2}\right)^2-\left(\omega
r_h\right)^2}-\frac{2}{3}p\\
b & =\alpha +\beta +\frac{2}{3}p=-\frac{i \omega r_h
}{3}+\frac{1}{2}-\frac{1}{3}\sqrt{\left(\lambda
_3+\frac{3}{2}\right)^2-\left(\omega
r_h\right)^2}+\frac{2}{3}p,\\
c & = 1+2\alpha =1+ \frac{2}{3}i \omega r_h
\end{aligned}
\end{equation}
% Choosing $\beta _-$ because the hypergeometric function converges
% only for $\Re(c-a-b)>0$, and taking the limit $r\to r_h$, we get:
% \begin{equation}\label{eq20}
% \begin{aligned}
%  \mathop {\lim }\limits_{\scriptstyle r \to r_h  \hfill \atop
%   \scriptstyle f \to 0 \hfill} R_{{\rm{NH}}} (f) & \simeq A_ -  f^{ - i\frac{{r_h \omega }}{3}} \left( {\frac{{r_h }}{r}} \right)^{\beta (n + 1)}  \times 1 + A_ +  f^{i\frac{{r_h \omega }}{3}} \left( {\frac{{r_h }}{r}} \right)^{\beta (n + 1)}  \times 1 \\
%   & = \left( {\frac{{r_h }}{r}} \right)^{\beta (n + 1)} \left[ {A_ -  e^{ - ir_h^4 \omega y}  + A_ +  e^{ir_h^4 \omega y} } \right] \\
%  \end{aligned}
% \end{equation}
% where $y \equiv \frac{{\ln f}}{{r_h^{n + 1} (n + 1)}}$. Now,
Imposing the boundary condition that at the horizon the solution be
purely in-going, we obtain the constraint
\begin{equation}\label{eq18a} A_+=0
\end{equation}
% so that near the horizon, only the
%$A_-$ solution survives:
%\begin{equation}\label{eq21}
%R_{\text{NH}}(f)=A_-f^{\alpha _{\pm }}(1-f){}^{\beta _-}F(a,b,c;f)
%\end{equation}
Far from the horizon ($f\to 1$), the asymptotic solution (\ref{eq18}) under the constraint (\ref{eq18a}) behaves as
\begin{eqnarray}\label{eq25}
 R_{{\mathrm{NH}}} (f) &\sim& A_ -  \left[ {\left( {\frac{{r_h }}{r}} \right)^{3\beta } \frac{{\Gamma (1 + 2\alpha )\Gamma (1 - 2\beta )}}{{\Gamma (1 + \alpha  - \beta  + {\textstyle{\frac{2}{3}}}p)\Gamma (1 + \alpha  - \beta  - {\textstyle{\frac{2}{3}}}p)}}}
\right. \nonumber\\
& & \left. {+ \left( {\frac{{r_h }}{r}} \right)^{3(1-\beta) } \frac{{\Gamma (1 + 2\alpha )\Gamma (2\beta  - 1)}}{{\Gamma (\alpha  + \beta  - {\textstyle{\frac{2}{3}}}p)\Gamma (\alpha  + \beta  + {\textstyle{\frac{2}{3}}}p)}}} \right]
\end{eqnarray}
which follows from standard identities of hypergeometric functions.

In the far field region, as $r\to \infty$, or correspondingly $f\to
1$, the radial wave equation may be approximated by
\begin{equation}\label{eq22}
g^{\prime\prime }(r)+\frac{1}{r}g'(r)+\left[\omega
^2-\left(\frac{9}{4}+\lambda _3\left(\lambda
_3+3\right)\right)\frac{1}{r^2}\right]g(r)=0
\end{equation}
where we defined $g(r)=r^{3/2} R(r)$.
The solution can be written in terms of Bessel functions,
%becomes a Bessel equation after the substitution
%$R(r)\equiv \frac{g(r)}{r^{3/2}}$:
%with the solution:
\begin{equation}\label{eq23}
R_{\text{FF}}(r)= \frac{g(r)}{r^{3/2}} = \frac{1}{r^{3/2}}\left(B_+J_{\lambda _3+3/2}(\omega r
 )+B_-Y_{\lambda _3+3/2}(\omega r)\right)
\end{equation}
In the low frequency regime,
\begin{equation}\label{eq23a}
\omega r_h \ll 1 \end{equation}
the asymptotic solution (\ref{eq23}) may be approximated near the horizon by
\begin{equation}\label{eq26}
R_{\text{FF}}(r)\sim B_+\frac{\left(\frac{\omega
}{2}\right)^{\lambda _3+3/2}}{\Gamma \left(\lambda
_3+5/2\right)}r^{\lambda _3} -B_-\frac{\Gamma \left(\lambda _3+3/2\right)}{\pi
}\left(\frac{\omega }{2}\right)^{-\lambda
_3-3/2}r^{-\lambda _3-3}
\end{equation}
Matching the two asymptotic solutions in the intermediate region ((\ref{eq25}) and (\ref{eq26}), respectively), we obtain
\begin{equation}\label{eq27}
\frac{B_+}{B_-}=-\frac{\Gamma (1-2\beta )\Gamma \left(\alpha +\beta
-\frac{2}{3}p\right)\Gamma \left(\alpha +\beta
+\frac{2}{3}p\right)\Gamma \left(\lambda _3+5/2\right)\Gamma
\left(\lambda _3+3/2\right)}{\pi  \Gamma \left(1+\alpha -\beta
+\frac{2}{3}p\right)\Gamma \left(1+\alpha -\beta
-\frac{2}{3}p\right)\Gamma (2\beta -1)}\left(\frac{2}{\omega
r_h}\right)^{2\lambda _3+3}
\end{equation}
where we used $\beta \approx -\lambda_3/3$ (eq.~(\ref{eq19})), on account of (\ref{eq23a}).

The reflection coefficient is
\begin{equation}\label{eq28}
\mathcal{R} =\frac{\text{outgoing amplitude}}{\text{incoming
amplitude}}=\frac{B_+-i
B_-}{B_++i B_-}
\end{equation}
from which we deduce the bulk absorption probability
\begin{equation}\label{eq28a}
\left|\mathcal{A}_p^{\mathrm{bulk}}\right|^2 =
1-\left|\mathcal{R}\right|^2 = \frac{2i \left(B^*-B\right)}{B
B^*+i\left(B^*-B\right)+1} \ \ , \ \ \ \ B = \frac{B_+}{B_-}
\end{equation}
After some
algebraic manipulations, in the low frequency regime (\ref{eq23a}) we arrive at the explicit expression for the absorption probability,
\begin{equation}\label{eq29}
\left|\mathcal{A}_p^{\mathrm{bulk}}\right|^2=4\pi \left(\frac{\omega
r_h}{2}\right)^{4+2\lambda _3}\left[ \frac{\Gamma
\left(1+\frac{\lambda _3+2p}{3}\right)\Gamma \left(1+\frac{\lambda
_3-2p}{3}\right)}{\Gamma \left(1+\frac{2\lambda _3}{3}\right)\Gamma
\left(\lambda _3+\frac{5}{2}\right)} \right]^2
\end{equation}
which reduces to the tensionless result in the limit $b\to 1$ (see, e.g.,
\cite{bibKantiI,bibCCG2,bibCEKT}).

\begin{figure}
  \includegraphics[width=6in]{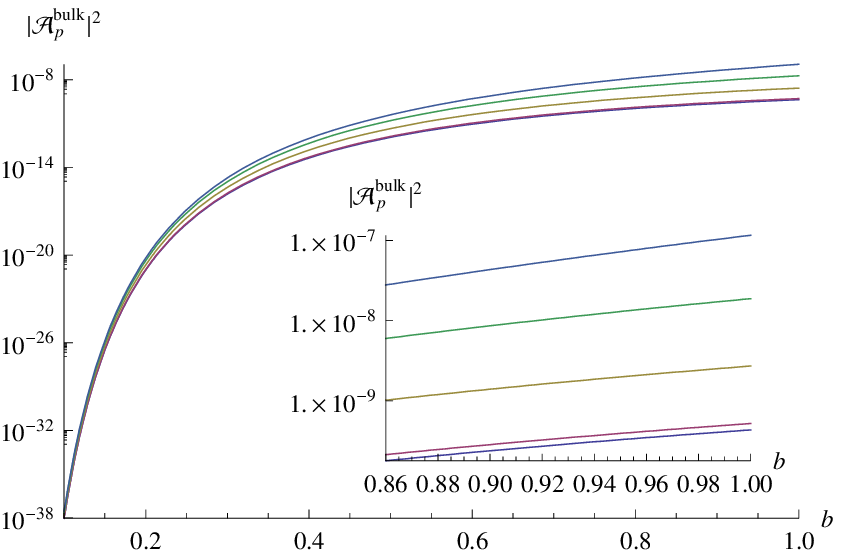}\\
  \caption{Low frequency absorption probability for bulk emissions as a function of $b$. Parameter $p\in
  \left\{0,\frac{1}{2},\frac{3}{2},2,2.25\right\}$ from eq. (\ref{eq13}) increases from
  bottom to top. Here, $\ell _3=2$, $m=1$, $\omega =0.2$, and $\mu =1$.}\label{fig1}
\end{figure}

\begin{figure}
  \includegraphics[width=6in]{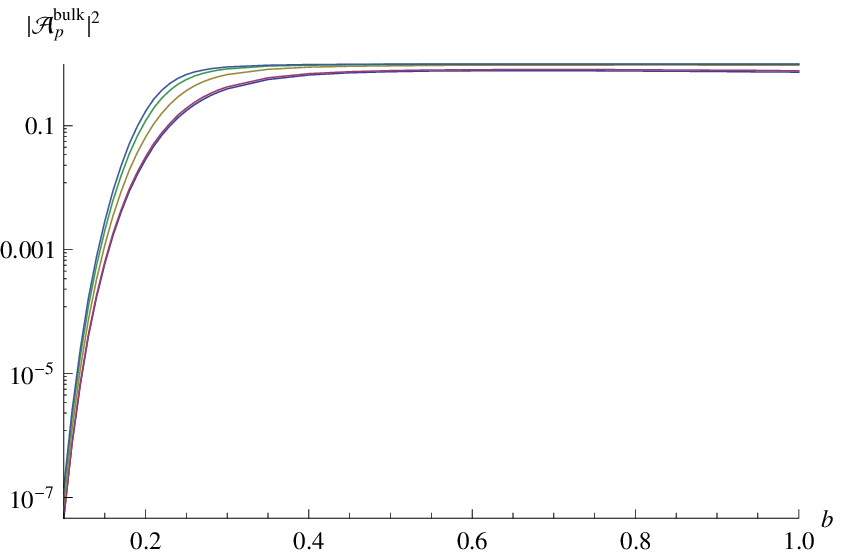}\\
  \caption{High frequency absorption probability for bulk emissions as a function of $b$. Parameter $p\in
  \left\{0,\frac{1}{2},\frac{3}{2},2,2.25\right\}$ from eq. (\ref{eq13}) increases from
  bottom to top. Here, $\ell _3=2$, $m=1$, $\omega =2.2$, and $\mu =1$.}\label{fig2}
\end{figure}

\begin{figure}
  \includegraphics[width=6in]{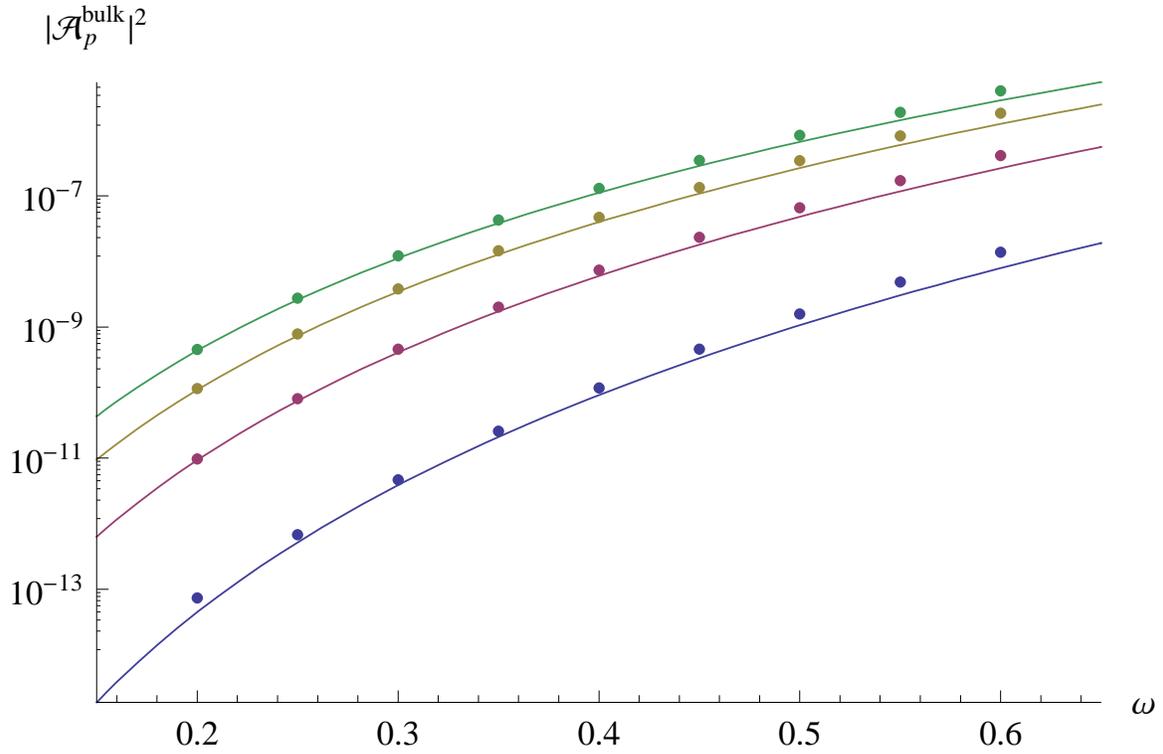}\\
  \caption{Absorption probability for bulk scalars ($p=0$) as a function of $\omega$ for various values of brane tension, where $b\in
  \{0.4,0.6,0.8,1\}$ increasing from bottom to top, with $\ell _3=2, m=1, \mu
  =1$. The points on the graph come from exact numerical calculations.}\label{fig3}
\end{figure}

\begin{figure}
  \includegraphics[width=6in]{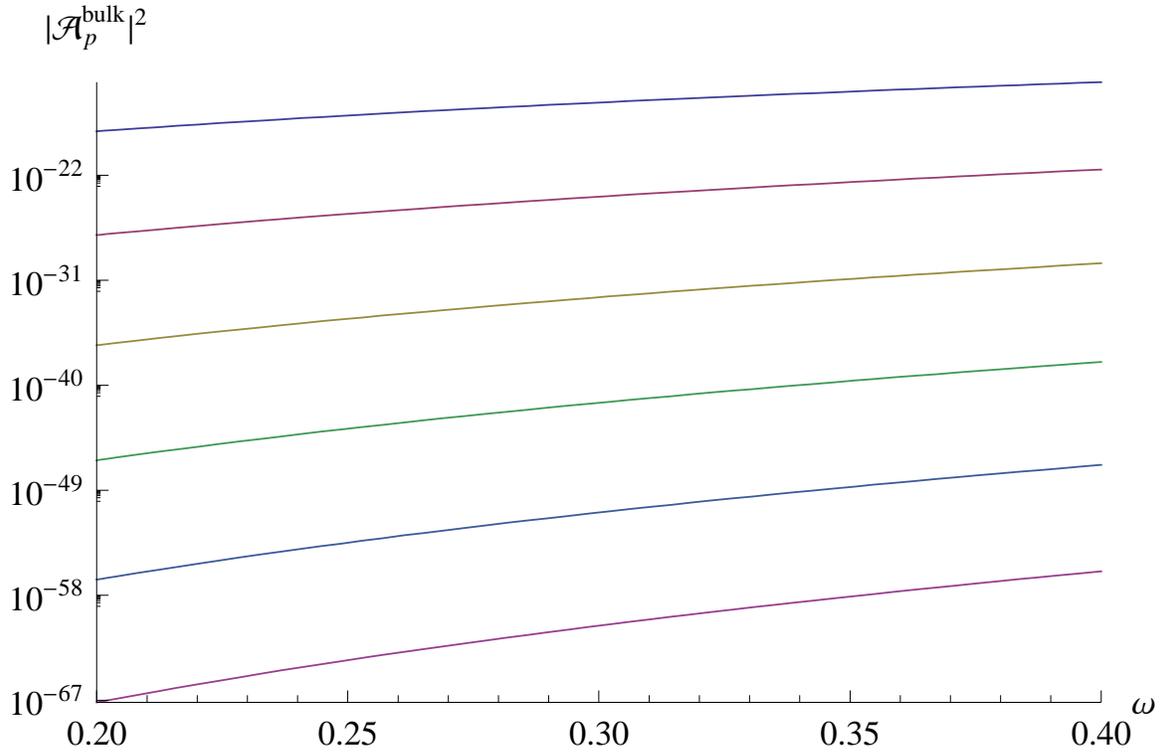}\\
  \caption{Bulk absorption probability for scalars for different values of
  $m$: for $\ell _3=5$, and $m\leq \ell _3$, $m$ increasing top to bottom, where
  $b=0.3$, $\mu =1$.}\label{fig4}
\end{figure}

Evidently, the dependence of the absorption probability on the brane
tension is complicated. To study its behavior, let us first look at
some graphical representations of this result. Figure \ref{fig1}
depicts the absorption probability as a function of $b$ with
$\omega$ fixed. As can be seen, all types of perturbation exhibit
the same qualitative behavior with the probability vanishing as
$b\to 0$ and increasing monotonically as $b\to 1$. This behavior
persists qualitatively at high frequency as demonstrated in figure
\ref{fig2} which was obtained from points generated by exactly
solving for absorption probability numerically. A further
demonstration of how tension suppresses the absorption probability
is provided in figure \ref{fig3}, where probability {\em vs}
frequency in the case of a scalar perturbation is plotted for
various values of $b$ in the small frequency regime. The graph also
has points representing exact numerical calculations. As expected,
the predictions of the analytical formula diverge from the exact
values with increasing frequency. Finally, figure \ref{fig4} shows
the effect of the magnetic quantum number $m$ in the case of scalar
emission. The qualitative behavior of all curves is the same,
however the absorption probability varies over many orders of
magnitude depending on $m$. In figs. \ref{fig3} and \ref{fig4},
%these last two graphs,
the type of perturbation chosen has a negligible effect on the
outcome.

% The effect that brane tension has on attenuating the spectrum can be
% seen in the grey-body spectra found numerically in \cite{bibDKSS}.
% To have a better idea of what is going on, and to be able to compare
% with brane-localized modes later on, we shall utilize the fact that
% $0<b\leq 1$, and expand the expression in (\ref{eq29}) around $b=0$,
% keeping only the lowest order term. The outcome of doing so looks
% like this:
% \begin{equation}\label{eq80}
% \left|\mathcal{A}_p\right|^2\sim \frac{\pi }{3}\left(\mu
% ^{1/3}\omega \right)^{2\left( \lambda
% _3 + 2\right)}\frac{b^{\frac{4 \lambda
% _3+5}{3}}}{2^{\frac{10\lambda
% _3+9}{3}}m^{2\lambda _3+3}}e^{\frac{2m}{b}},\qquad
% m\neq 0
% \end{equation}
% where the expression was derived starting from the general
% higher-dimensional expression in the tenseless case to explicitly
% show the structure of the numerical factors appearing in it; it
% being understood that only $n=2$ is relevant here. The expression
% for $m=0$ can easily be obtained from (\ref{eq29}) without
% expansion. We can still see $b$ lurking inside the $\lambda_3$s, but
% the dependence on $b$ is now more explicit, and the way the
% expression tends to zero as $b\to 0$ is easily shown by rewriting
% the whole thing as a single exponential, and by inspecting the
% exponent we see it involves either terms of the form $-1/b$ or $(x+y
% \ln b)/b$ (for positive $y$) both leading to zero when exponentiated
% as we take the limit $b\to 0$. In the $m=0$ case, the same
% considerations apply.

Moving on to the case of brane localized modes (the photon, massless
fermion and scalar), we note that because we are projecting onto
the brane, angles $\chi$ and $\psi$ will be fixed, so that the
dependence of the metric on $b$ will only be through $r_h$ inside
the metric function $f$. Aside from that, the results pertaining to the
grey-body factors will be identical to those in the tensionless
case. To find these grey-body factors we follow the same procedure
as above, but instead using an equation analogous to the Teukolsky
equation on the brane. For a static background, the radial wave
equation is (see, e.g., \cite{bibCL,bibKantiII})
\begin{equation}\label{eq36}
\begin{aligned}
  (fr^2 )^s & \frac{{\text{d}}}
{{{\text{d}}r}} \left[ {(fr^2 )^{1 - s} \frac{{{\text{d}}R_s }}
{{{\text{d}}r}}} \right] + \left\{ {\frac{{\omega ^2 r^2 }} {f} +
2is\omega r - \frac{{is\omega r}}
{f}(n + 1)(1 - f)} \right. \hfill \\
  & \left. { - \Lambda  - (2s - 1)(s - 1)(n + 1)(1 - f)} \right\}R_s  = 0 \hfill \\
\end{aligned}
\end{equation}
where $\Lambda  \equiv \ell (\ell  + 1) - s(s - 1)$, and $s$ is the
spin. The corresponding angular equation reads
\begin{equation}\label{eq37}
\frac{1} {{\sin \theta }}\frac{{\text{d}}} {{{\text{d}}\theta
}}\left( {\sin \theta \frac{{{\text{d}}{}_sS_\ell ^m }}
{{{\text{d}}\theta }}} \right) + \left( { - \frac{{2ms\cot \theta }}
{{\sin \theta }} - \frac{{m^2 }} {{\sin ^2 \theta }} + s - s^2 \cot
^2 \theta  + \Lambda } \right){}_sS_\ell ^m  = 0
\end{equation}
where $e^{i m \varphi } \, _sS_{\ell }^m(\theta )= \, _sY_{\ell
}^m\left(\Omega _2\right)$ are known as the spin-weighted spherical
harmonics. For $s=\frac{1}{2}$, the absorption probability is
\cite{bibKantiII}:
%Because of this, calculating the grey-body factors in the present
%case follows derivations in
%closely, except for the modifications prescribed in (\ref{eq5}) to
%take tension into account, and they are essentially the same as
%those followed for bulk emissions above, and so we shall only list
%the final results directly, more details can be found in the
%above-cited papers. In the limit where $\omega r_h \ll 1$, we have
%\cite{bibKantiII}
\begin{equation}\label{eq38}
\left|\mathcal{A}_{s=1/2}^{\mathrm{brane}}\right|^2= 4\pi \left(
\frac{\omega r_h}{2^{5/3}} \right)^{2j+1} \frac{1}{[\Gamma
\left(j+1\right)]^2}
\end{equation}
and for $s=1$
\begin{equation}\label{eq39}
\left|\mathcal{A}_{s=1}^{\mathrm{brane}}\right|^2=\frac{1}{9}
\left(2\omega r_h\right)^{2j+2} \left[ \frac{\Gamma
\left(\frac{j}{3}\right)\Gamma \left(\frac{j+1}{3}\right)\Gamma
\left(j+2\right)}{\Gamma \left(\frac{2j+1}{3}\right)\Gamma
\left(2j+2\right)}\right]^2
\end{equation}
where $j$ is the total angular momentum in both cases.

\begin{figure}
  \includegraphics[width=6in]{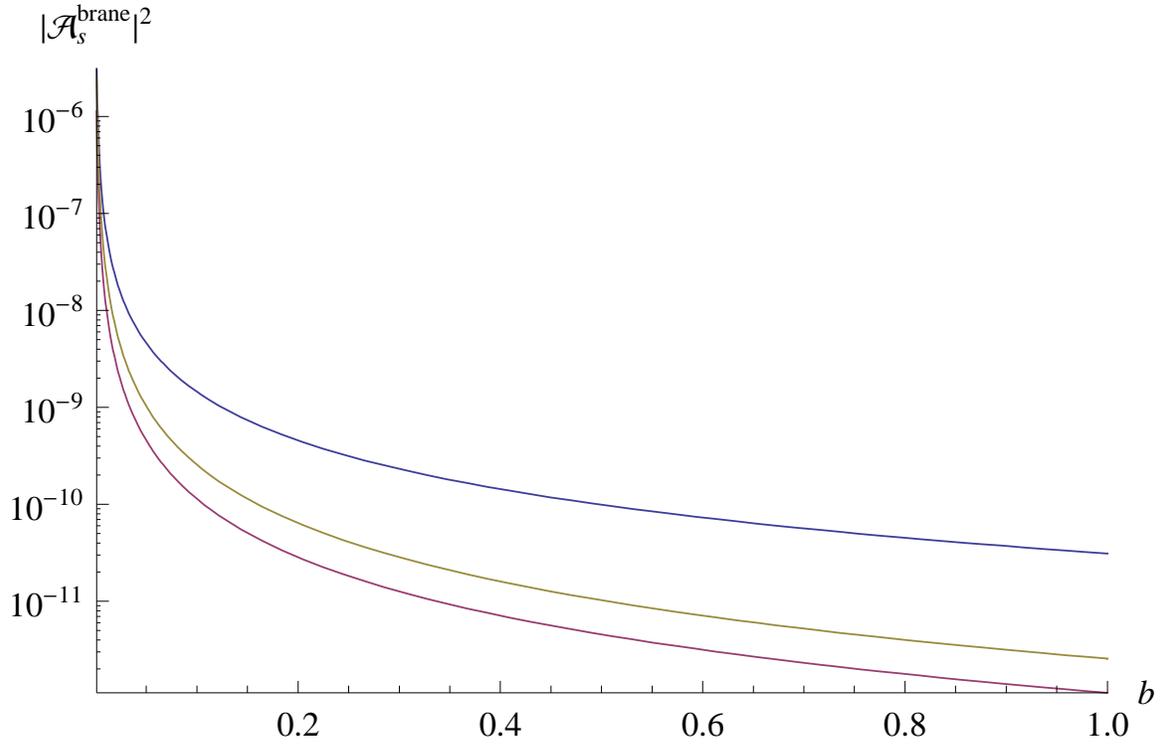}\\
  \caption{Absorption probability for brane-localized emissions as a function of $b$. The curves from top to bottom correspond to
  $s=\frac{1}{2}$, $s=0$ and $s=1$, respectively. Here, $\ell=2$,
  $\omega =0.02$, and $\mu =1$.}\label{fig5}
\end{figure}

\begin{figure}
  \includegraphics[width=6in]{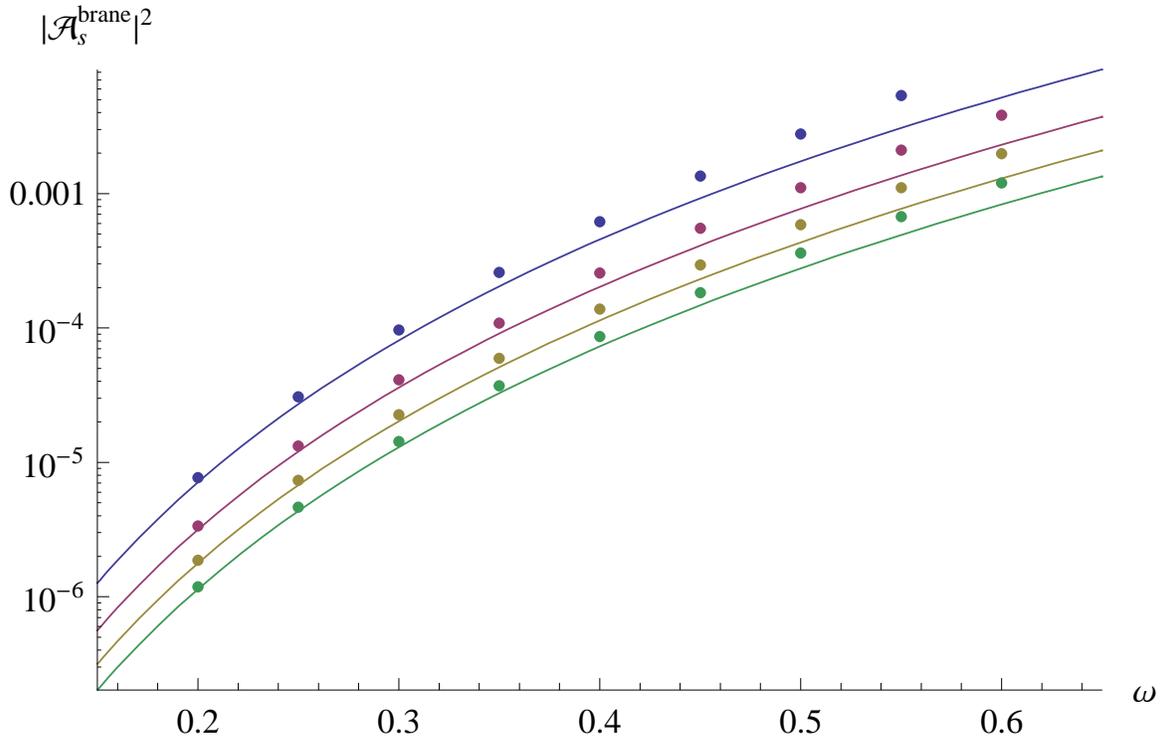}\\
  \caption{Absorption probability for brane-localized scalars ($s=0$) as a function of $\omega$ for various values of brane tension, where $b\in
  \{0.4,0.6,0.8,1\}$ increasing from top to bottom, with $\ell=2, m=1, \mu
  =1$. The points on the graph come from exact numerical calculations.}\label{fig6}
\end{figure}

For brane-localized scalars the result is \cite{bibKantiI}:
\begin{equation}\label{eq79}
\left| {{\mathcal A}_{s = 0}^{\mathrm{brane}} } \right|^2  =
\frac{{16\pi }} {9}\left( {\frac{{\omega r_h }} {2}} \right)^{2j  +
2} \left[ {\frac{{\Gamma \left( {\frac{{j  + 1}} {3}} \right)\Gamma
\left( {1 + \frac{{j }} {3}} \right)}} {{\Gamma \left( {\frac{1} {2}
+ j } \right)\Gamma \left( {1 + \frac{{2j  + 1}} {3}} \right)}}}
\right]^2
\end{equation}
Upon comparison of (\ref{eq39}) and (\ref{eq79}), we obtain
\begin{equation}\label{eq81}
\frac{\left|\mathcal{A}_{s=0}^{\mathrm{brane}}\right|^2}{\left|\mathcal{A}_{s=1}^{\mathrm{brane}}\right|^2}=\left(\frac{j}{1+j}\right)^2
\end{equation}
which shows that (in this frequency regime) the gauge vector and
scalar behave the same way up to a multiplicative factor, which tends to
unity for large $j$. One can see from the expressions above that
only the first few values of $j$ contribute to the emission spectrum
making the emission rate for scalars higher than for gauge vectors
in this frequency range. This can be seen numerically in the
absence of tension in the results of \cite{bibJungPark}. At higher
frequencies the situation reverses, and the rate for scalars becomes
smaller. Graphs analogous to those for bulk emissions are shown in
figures \ref{fig5} and \ref{fig6} for scalars. For low frequencies,
the behavior with varying tension is now reversed and emission is
enhanced with increased tension. For higher frequencies the trend is
not as conclusive. The difference between the two cases can be
understood in terms of two competing effects: {\em (a)} the enhancement of
the geometrical cross-section with tension implies an increased rate of
emission for both bulk and brane-localized modes, and {\em (b)} the fact that
the angular eigenvalues increase with tension and appear
only in the bulk potential (causing the potential barrier to
increase) results in suppression of only bulk
emissions.

From an experimental point of view, it is of great interest to know
the relative emissivities between brane and bulk modes. The above
discussion seems to hint that the inclusion of tension favors the
EHM conjecture \cite{bibbh3}, namely, that black holes emit mainly
on the brane. To verify this in our case, let us consider the
emission spectrum as defined in eq. (\ref{eq1}) for scalars. For
that, we need to write the grey-body factor $\sigma _{\ell ,\kappa}
(\omega )$ in terms of the absorption probability $|{\cal A}|^2$ and
scalar multiplicities $N_{\ell,m}$. For either bulk or
brane-localized emissions it is of the form
\begin{equation}\label{eq82}
\frac{{{\text{d}}E(\omega )}} {{{\text{d}}t\,{\text{d}}\omega }} =
\frac{1} {{2\pi }}\sum\limits_{\ell ,m} {N_{\ell ,m} \left| {{\cal
A}_{\ell ,m} } \right|^2 \frac{\omega } {{e^{\omega /T_{\text{H}} }
- 1}}}
\end{equation}
In the bulk, the scalar multiplicities are \cite{bibDKSS}
\begin{equation}\label{eq83}
N_{\ell _3 ,m}^{{\text{scal}}}  = \left\{ {\begin{array}{*{20}c}
   {m = 0,} & {\frac{1}
{2}\left( {\ell _3  + 2} \right)\left( {\ell _3  + 1} \right)}  \\
   {m \ne 0,} & {\left( {\ell _3  - m + 2} \right)\left( {\ell _3  - m + 1} \right)}  \\
 \end{array} } \right.
\end{equation}
whereas on the brane they they are equal to the standard value $2\ell+1$. Starting with
the bulk, we numerically solve eq.~(\ref{eq16}) for the absorption
probability with $p=0$, and using eq.~(\ref{eq83}), we end up with
the spectrum shown in fig. \ref{fig8}. A similar calculation for brane-localized scalars is done by projecting the bulk
equation on the brane, and the result is shown in fig.~\ref{fig9}.
Notice that in both figures, the emission rates are enhanced
with tension. The reason for this comes primarily not from the
grey-body factor, but rather from the denominator in the black-body
distribution which depends on the Hawking temperature that is
common to both types of emission. We can discount this effect by
looking at the relative emissivity, in which we take the ratio of
the emission rate in the bulk to that on the brane. Figure
\ref{fig10} shows this for scalars.

Similar plots are obtained for gauge vectors. On the brane, we again
have the multiplicities $2\ell+1$, but in the bulk we have
\begin{equation}\label{eq84}
N_{\ell _3 ,m}^{{\text{vec}}}  = \left\{ {\begin{array}{*{20}c}
   {m = 0,} & {\frac{1}
{2}\ell _3 \left( {3\ell _3  + 5} \right)}  \\
   {m \ne 0,} & {3\left( {\ell _3  - m + 2} \right)\left( {\ell _3  - m + 1} \right)}  \\

 \end{array} } \right.
\end{equation}
In the bulk, we solve eq.~(\ref{eq16}) with $p=\frac{1}{2}$, and on
the brane eq. (\ref{eq36}) is solved with $s=1$. The results of
doing this are shown in figures \ref{fig11} through \ref{fig13}. In
this case the emissivity is only partial, since the scalar degree of
freedom is not included.

We can see that in both cases, in the range shown, the emission rate
is greater on the brane than in the bulk, in support of the EHM
claim. It should be pointed out, however, that even though mini
black-holes at the LHC will spend most of their life in the
Schwarzschild phase, they first form with
non-zero angular momentum. This has been shown to modify the
Hawking radiation due to super-radiance in such a way that emission
in the bulk at the initial phase dominates over emission on the brane
\cite{bibFS1,bibFS2,bibSt}.

\section{High-frequency grey-body factors}
\label{sec4}

Next we turn our attention to grey-body factors at large imaginary
frequencies. We shall discuss different values of
spin and take into account the extra dimensions as
well as the brane tension in the codimension-2 model \cite{bibKK}. The
calculation will follow the method of Andersson and Howls
\cite{bibAndHowls} which was then used by Cho \cite{bibCho1}.
It is a combination of the monodromy argument of Motl and Neitzke
\cite{bibMotlN} and the standard complex-coordinate WKB method.

The first step is to rewrite the master equation (\ref{eq11}) in a form suitable
for the application of the WKB method. To this end, we introduce the redefinition of the wavefunction,
\begin{equation}\label{eq30}
\Psi =\frac{\phi }{\sqrt{f}}
\end{equation}
which reduces (\ref{eq11}) to
\begin{equation}\label{eq31}
\phi''(r)+Q^2(r)\phi (r)=0
\end{equation}
The resulting WKB solution is
\begin{equation}\label{eq32}
\phi _{1,2}^{(t)}=\frac{1}{\sqrt{Q(r)}}\exp \left[\pm i \int_t^r
Q\left(r'\right) \, dr'\right]
\end{equation}
with $t$ being a reference point, and
%where, for the appropriate
%potential we have:
\begin{equation}\label{eq32a}
Q^2(r)=\frac{1}{f^2(r)}\left[\omega
^2-V(r)+\frac{1}{4}(f'(r))^2-\frac{1}{2}f(r)f^{\prime\prime
}(r)\right]
\end{equation}
Integer spin particle are described by the potential (\ref{eq9}).

Following Motl and Neitzke \cite{bibMotlN}, we go around a
contour in the complex $r$-plane and impose preservation of
monodromy. Because of the exponential nature of the WKB solutions,
dominance will be exchanged between the two exponentials as we move
around the complex plane. This implies that some terms that are
exponentially small and can be overlooked in one region of the
complex plane can grow exponentially in a different region. To take this
into account, one needs to resort to the Stokes phenomenon which
keeps track of these changes.

\begin{figure}
  \includegraphics[width=4in]{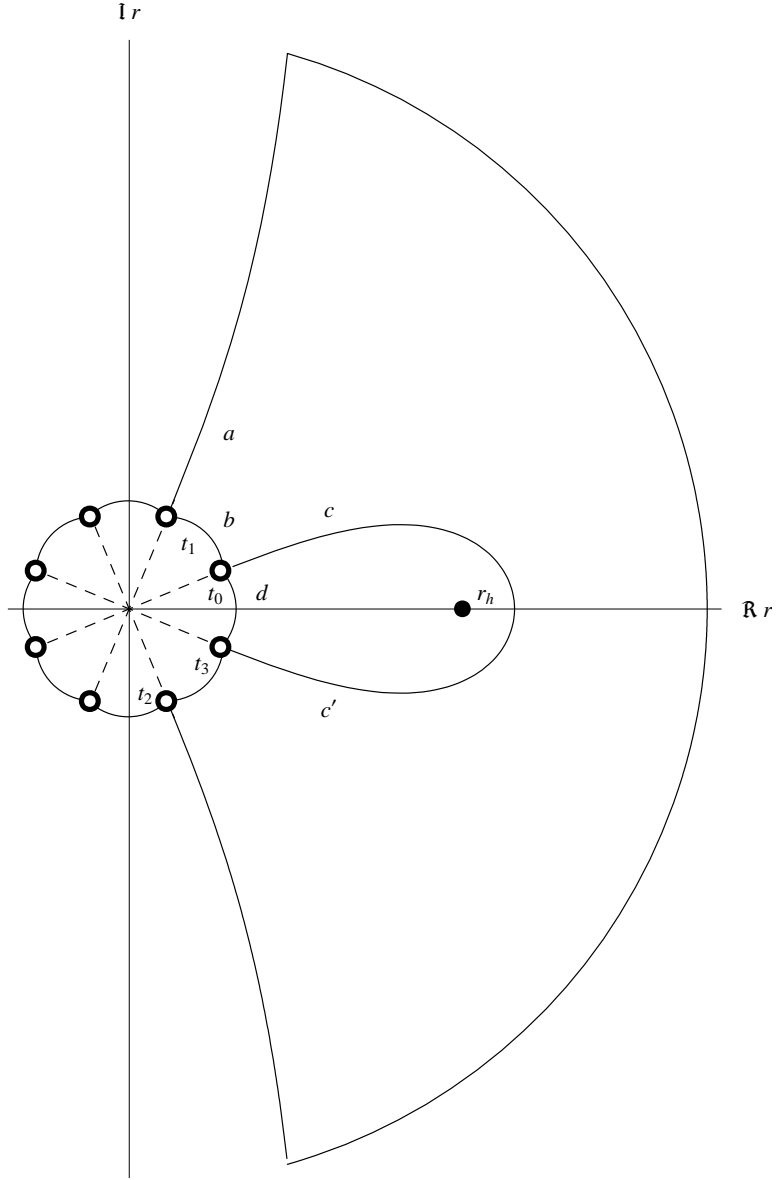}\\
  \caption{Stokes lines structure for the Schwarzschild black hole in 6D in the complex $r$-plane for large imaginary frequency. Empty circles are the zeros of $Q$ and the filled circle is the horizon radius.}\label{fig7}
\end{figure}

\begin{figure}
  \includegraphics[width=6in]{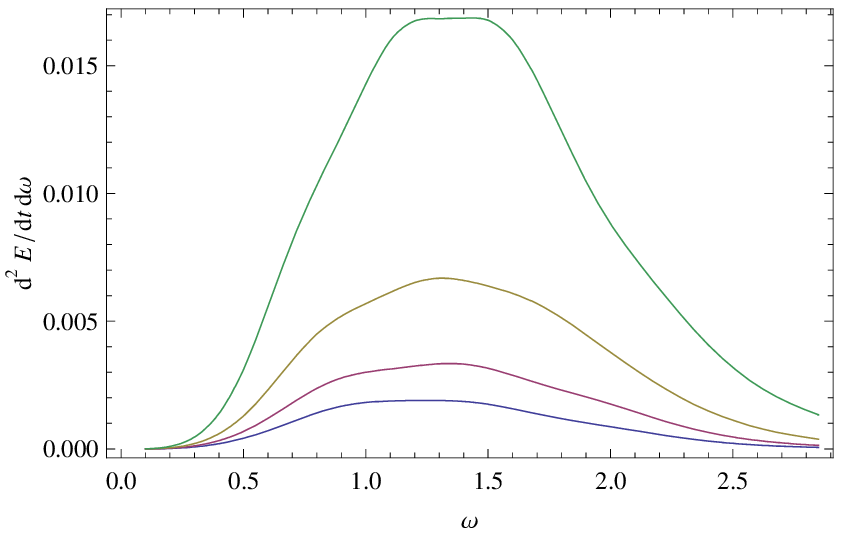}\\
  \caption{Energy emission spectrum for scalars in the bulk for various values of brane tension, with $b\in\{0.4,0.6,0.8,1\}$ increasing from top to bottom.}\label{fig8}
\end{figure}

\begin{figure}
  \includegraphics[width=6in]{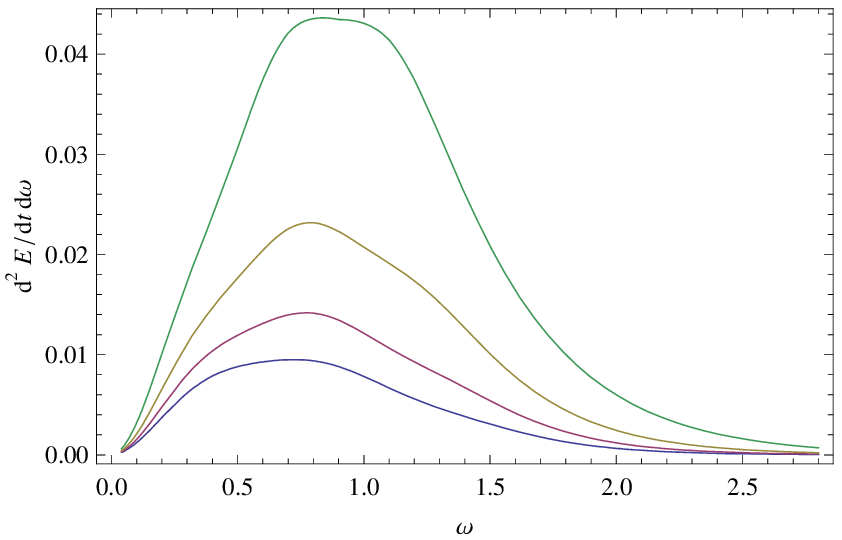}\\
  \caption{Energy emission spectrum for scalars on the brane for various values of brane tension, with $b\in\{0.4,0.6,0.8,1\}$ increasing from top to bottom.}\label{fig9}
\end{figure}

\begin{figure}
  \includegraphics[width=6in]{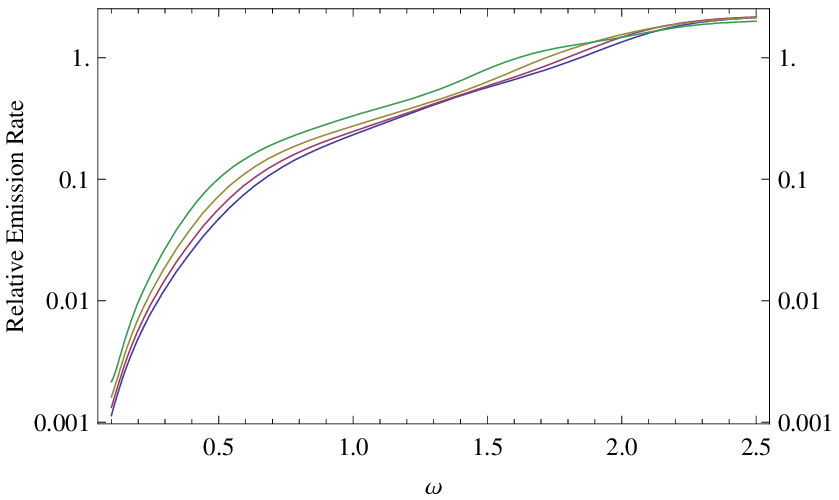}\\
  \caption{Relative bulk-to-brane emissivity for scalars for different values of brane tension, with $b\in\{0.4,0.6,0.8,1\}$ increasing from top to bottom.}\label{fig10}
\end{figure}

\begin{figure}
  \includegraphics[width=6in]{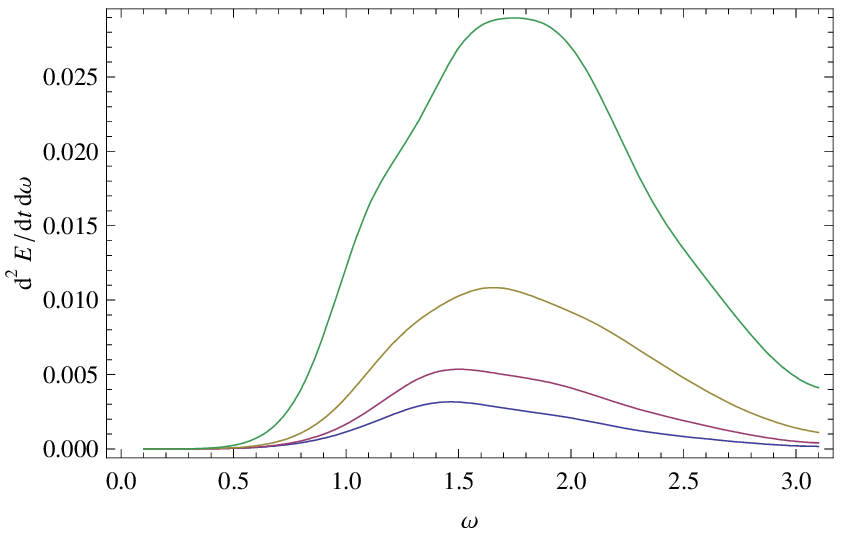}\\
  \caption{Energy emission spectrum for gauge vectors in the bulk for various values of brane tension, with $b\in\{0.4,0.6,0.8,1\}$ increasing from top to bottom.}\label{fig11}
\end{figure}

\begin{figure}
  \includegraphics[width=6in]{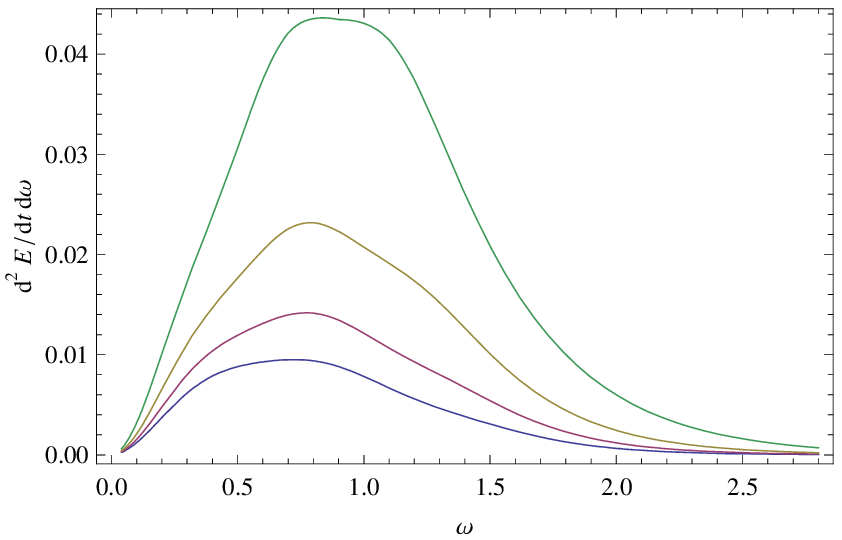}\\
  \caption{Energy emission spectrum for gauge vectors on the brane for various values of brane tension, with $b\in\{0.4,0.6,0.8,1\}$ increasing from top to bottom.}\label{fig12}
\end{figure}

\begin{figure}
  \includegraphics[width=6in]{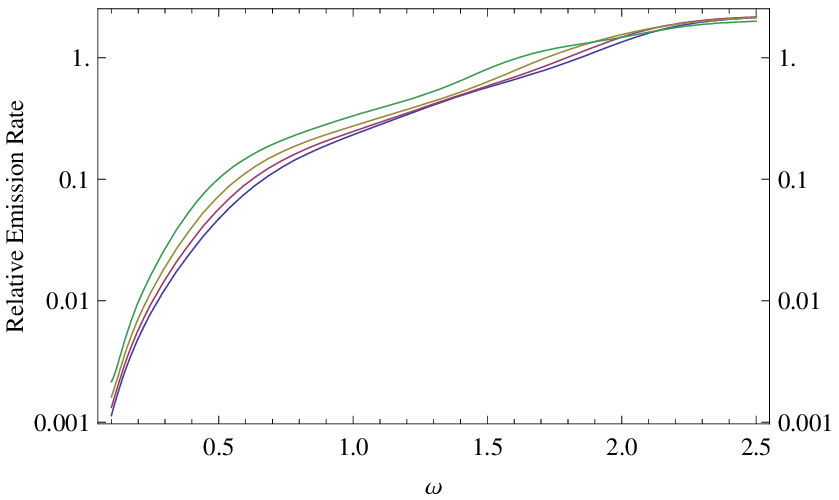}\\
  \caption{Relative bulk-to-brane emissivity for gauge vectors for different values of brane tension, with $b\in\{0.4,0.6,0.8,1\}$ increasing from top to bottom.}\label{fig13}
\end{figure}

In such an analysis, we need to pay attention to the zeroes and
poles of $Q(r)$. In particular, it can be shown that from each
simple zero of $Q$ there emanate 3 Stokes lines along which
$Q(r)\mathrm{d} r$ is imaginary. This means that one of the two
solutions $\phi _{1,2}^{(t)}$ will grow exponentially whereas the
other one will decay as we move away from the reference point $t$.
On the other hand, we can also define anti-Stokes lines emanating
from each zero of $Q$ along which $Q(r) \mathrm{d} r$ is real and
$\phi _{1,2}^{(t)}$ are oscillatory functions. By crossing
anti-Stokes lines, the exponential behaviors of the two solutions
are switched, while extending the solution across a Stokes line, the
linear combination defining the solution is changed in a
well-defined way: the coefficient of the dominant term stays the
same, but that of the sub-dominant term picks up a contribution
proportional to the coefficient of the dominant term. Thus we see
that the appropriate contour that can be used will trace anti-Stokes
lines and cross Stokes lines. Figure \ref{fig7} shows such a contour
for Schwarzschild geometry assuming (large) purely imaginary
frequency.

One can see from the potential (\ref{eq9}) that for $|\Im \omega|
\to \infty$, all zeroes of $Q$ approach the origin and the potential
may be approximated by its expansion around $r=0$. In this case, the
$\ell$-dependent terms are negligible. They enter in subleading
contributions \cite{bibMuSi,bibAlec}. Near $r=0$, keeping only the
most singular term, the potential in (\ref{eq9}), which describes
integer-spin perturbations, becomes:
\begin{equation}\label{eq72}
V_{{\text{bulk}}}  \sim \frac{1} {4}(2 + n)^2 (p^2  -
1)\frac{{r_h^{2(1 + n)} }} {{r^{2(2 + n)} }}
\end{equation}
and the resulting WKB equation is:
\begin{equation}\label{eq33}
\psi ^{\prime\prime }(r)+R_0(r)\psi (r)\simeq 0
\end{equation}
where
\begin{equation}\label{eq34}
R_0(r)=\left(\frac{r}{r_h}\right){}^{2n+2}\left[\omega
^2-\frac{1}{4r^{2(n+2)}}\left(\left(p^2-1\right)(n+2)^2+(n+1)(n+3)\right)r_h^{2(n+1)}\right]
\end{equation}
However, one can show \cite{bibAndHowls} that the solutions $\phi
_{1,2}^{(t)}$ so obtained by identifying $R_0(r)$ with $Q^2(r)$, do
not have the right behavior near the origin when compared with the
exact solution near there. This can be rectified by using a slightly
different $R(r)$, which is permissible. The modified form turns out
to be:
\begin{equation}\label{eq35}
R(r)=R_0(r)-\frac{1}{4r^2}\equiv
Q^2(r)=\left(\frac{r}{r_h}\right){}^{2(n+1)}\left[\omega
^2-\frac{(n+2)^2p^2}{4r^{2(n+2)}}r_h^{2(n+1)}\right]
\end{equation}
For $\omega =-i|\omega |$, the zeros of $Q^2$ are:
\begin{equation}\label{eq40}
r_k=\left(\frac{(n+2)p}{2|\omega
|}\right)^{\frac{1}{(n+2)}}r_h^{\frac{n+1}{n+2}}e^{\frac{\pi
i}{n+2}\left(k+\frac{1}{2}\right)},\quad k=0,1,\cdots 2(2+n)-1
\end{equation}
These zeros will serve as reference points for the phase integrals,
because in doing so, the Stokes constants are either $+i$ for
traveling anti-clockwise (or $-i$ for clockwise), and so, we will
need to switch between reference points as we go around the contour:
\begin{equation}\label{eq47}
\phi _{1,2}^{\left(t_k\right)}=e^{\mp i \nu _{j k}}\phi
_{1,2}^{\left(t_j\right)}
\end{equation}
with
\begin{equation}\label{eq41}
\nu _{j k}=\int_{r_j}^{r_k} Q \, \mathrm{d}r=\frac{\omega
}{r_h^{n+1}}\int_{r_j}^{r_k}
r^{n+1}\left[1-\frac{(n+2)^2p^2}{4\omega
^2r^{2(n+2)}}r_h^{2(n+1)}\right]{}^{1/2} \, \mathrm{d}r
\end{equation}
where $j$ and $k$ label two consecutive zeros. Using the
substitution $y\equiv\frac{2\omega r^{n+2}}{(n+2)p r_h^{n+1}}$,
followed by $y\equiv \cosh x$, we get:
\begin{equation}\label{eq42}
\nu _{j k}=\int_{r_j}^{r_k} Q \, dr=\frac{p}{2}\int _{e^{i \pi
j}}^{e^{i \pi  k}}\left[1-\frac{1}{y^2}\right]^{1/2}d
y=\left(\frac{e^{i \pi  j}-e^{i \pi  k}}{2}\right)\frac{\pi  p}{2}
\end{equation}
so that:
\begin{equation}\label{eq46}
\nu_{10} = \nu_{30} \equiv \nu = -\frac{\pi p}{2}\equiv
\nu_{\mathrm{bulk}}
\end{equation}
It is also possible to consider a massless Dirac fermion propagating
in the bulk, which can be relevant in split-fermion theories where
rapid proton decay can suppressed by introducing bulk fermions
\cite{bibCCDN2,bibCCDN3} that are superpartners to localized scalars
and gauge bosons. A master equation potential based on a
higher-dimensional Dirac equation potential was derived in
\cite{bibCCDN3},
\begin{equation}\label{eq76}
V_{{\text{bulk-Dirac}}}  = f\frac{{\text{d}}} {{{\text{d}}r}}\left[
{\sqrt f \left( {\frac{{\ell  + \frac{{D - 2}} {2}}} {r}} \right)}
\right] + f\left( {\frac{{\ell  + \frac{{D - 2}} {2}}} {r}}
\right)^2
\end{equation}
By expanding near $r=0$, we find this potential to be proportional
to $r^{ - (7 + 3n)/2}$, which, for the present purpose, can be
treated as zero when compared to the other bulk perturbation given
by (\ref{eq72}). At this approximation, therefore, we may use $p=1$
for a bulk Dirac fermion.

Turning our attention to the actual calculation of grey-body
factors, we see that since $\omega$ in the present case is not real,
the conservation of flux will take on the following generalized
form:
\begin{equation}\label{eq44}
T(\omega)T(-\omega)+R(\omega)R(-\omega)=1
\end{equation}
and therefore, we consider two separate solutions representing the
boundary condition at physical infinity; one with frequency $\omega
$ and the other with $-\omega$. In the first case, the reflected
wave at infinity is represented by $\phi _1^{\left(t_1\right)}(r)$,
while the ingoing wave by $\phi _2^{\left(t_1\right)}(r)$. For
$-\omega$, the two solutions switch roles, but the calculation is
effectively the same, and we shall combine them into one by writing
the wavefunction in general as:
\begin{equation}\label{eq43}
\psi =X e^{i \theta }\phi _1^{\left(t_j\right)}+Y e^{-i \theta }\phi
_2^{\left(t_j\right)}
\end{equation}
where $\theta$ is a phase angle. In the case of $+\omega$, $X = R_+$
and $Y = 1$, whereas for $-\omega$, $X = 1$ and $Y = R_-$. For
$\Re(\omega  M)>0$, the outgoing wave boundary condition at spatial
infinity can be analytically continued to the anti-Stokes line
labeled $a$ in the figure. To find the the $T$'s and $R$'s we need
to travel a path on anti-Stokes lines starting and ending at point
$a$. The solution $\psi_a$ at point $a$ for $\pm\omega$ is described
by the wavefunction given in (\ref{eq43}) with $j=1$. Since we don't
cross any Stokes lines in extending the solution to $t_1$, the above
expression remains valid there, but in going to point $b$ we do
cross a Stokes line, and to account for the Stokes phenomenon we
need to add to $\psi_a$ a contribution proportional to $\phi_2$ with
the constant of proportionality equal to the Stokes constant
multiplied by the coefficient of the dominant function on the Stokes
line, $\phi_1$ in this case, so that:
\begin{equation}\label{eq45}
\psi _b=\psi _a-i X e^{i \theta }\phi _2^{\left(t_1\right)}=X e^{i
\theta }\phi _1^{\left(t_1\right)}+\left(Y e^{-i \theta }-i X e^{i
\theta }\right)\phi _2^{\left(t_1\right)}
\end{equation}
To extend the solution to point $c$, we now need to change the lower
limit of integration to have $t_0$ as reference:
\begin{equation}\label{eq48}
\psi _b=X e^{i \theta -i \nu }\phi _1^{\left(t_0\right)}+\left(Y
e^{-i \theta +i \nu }-i X e^{i \theta +i \nu }\right)\phi
_2^{\left(t_0\right)}
\end{equation}
In going to point $c$ there will be no exchange in dominance, since
we cross no anti-Stokes lines:
\begin{equation}\label{eq49}
\psi _c=\psi _b-i X e^{i \theta -i \nu }\phi _2^{\left(t_0\right)}=X
e^{i \theta -i \nu }\phi _1^{\left(t_0\right)}+\left(Y e^{-i \theta
+i \nu }-i X e^{i \theta +i \nu }-i X e^{i \theta -i \nu
}\right)\phi _2^{\left(t_0\right)}
\end{equation}
We now go to point $c'$, and in doing so we cross no Stokes lines,
so that the combination (\ref{eq49}) remains valid, but now the
integral is evaluated around a contour that loops around the pole
$r=r_h$. Replacing that contour by the one to the left of the pole,
we have:
\begin{equation}\label{eq50}
\psi _{c'}=X e^{i \theta -i \nu }e^{i \Gamma }\phi
_1^{\left(t_0\right)}+\left(Y e^{-i \theta +i \nu }-i X e^{i \theta
+i \nu }-i X e^{i \theta -i \nu }\right)\phi _2^{\left(t_0\right)}
e^{-i \Gamma }
\end{equation}
where $\Gamma$ is the integral encircling $r=r_h$ clockwise:
\begin{equation}\label{eq74}
\Gamma _{{\text{bulk}}}  = \mathop{\int\mkern-20.8mu
\circlearrowright}
 {Q\,{\text{d}}r}  =  - 2\pi i\mathop {\operatorname{Res} }\limits_{r = r_h } Q =  - i\pi \sqrt {1 + \left( {\frac{{2\omega r_h }}
{{n + 1}}} \right)^2 }  \approx  - \frac{1} {2}i\pi \beta
\end{equation}
where $\beta=1/T_{\mathrm{H}}$. To connect the solution to point
$d$, we need to change the lower integration limit to $t_3$:
\begin{equation}\label{eq51}
\psi _{c'}=X e^{i \theta }e^{i \Gamma }\phi
_1^{\left(t_3\right)}+\left(Y e^{-i \theta }-i X e^{i \theta -2i \nu
}-i X e^{i \theta }\right)\phi _2^{\left(t_3\right)} e^{-i \Gamma }
\end{equation}
and now to connect to point $d$, we need to cross the anti-Stokes
line to get inside the loop, and this means that $\phi _1$ and $\phi
_2$ will interchange dominance, further, the Stokes constant is now
$i$, since we have reversed direction:
\begin{equation}\label{eq52}
\begin{aligned}
\psi _d & =\psi _{c'}+i \left(Y e^{-i \theta }-i X e^{i \theta -2i
\nu }-i X e^{i \theta }\right)\phi _1^{\left(t_3\right)} e^{-i
\Gamma }\\
& =\left[X e^{i \theta }e^{i \Gamma }+\left(i Y e^{-i \theta }+ X
e^{i \theta -2i \nu }+X e^{i \theta }\right)e^{-i \Gamma }
\right]\phi _1^{\left(t_3\right)}\\
& +\left(Y e^{-i \theta }-i X e^{i \theta -2i \nu }-i X e^{i \theta
}\right)\phi
_2^{\left(t_3\right)} e^{-i \Gamma }\\
\end{aligned}
\end{equation}
Then switching back to point $t_0$:
\begin{equation}\label{eq53}
\begin{aligned}
\psi _d& =\left[X e^{i \theta }e^{i \Gamma }e^{-i \nu }+\left(i Y
e^{-i \theta }e^{-i \nu }+ X e^{i \theta -3i \nu }+X e^{i \theta
}e^{-i \nu }\right)e^{-i \Gamma } \right]\phi
_1^{\left(t_0\right)}\\
& +\left(Y e^{-i \theta }e^{i \nu }-i X e^{i \theta -i \nu }-i X
e^{i \theta }e^{i \nu }\right)\phi _2^{\left(t_0\right)} e^{-i
\Gamma }
\end{aligned}
\end{equation}
Going back to point $c$ and crossing a Stokes line, we encircle the
pole at the origin:
\begin{equation}\label{eq54}
\begin{aligned}
\psi _{\bar{c}}& =\left[\left(2e^{-i \nu }+e^{i \nu }+e^{-3i \nu
}\right)e^{i(\theta -\Gamma )}X+e^{i(-\nu +\Gamma +\theta )}X+i
\left(e^{-i \nu }+e^{i \nu }\right)e^{-i (\Gamma +\theta
)}Y\right]\phi _1^{\left(t_0\right)}\\
& +e^{-i \Gamma }\left[-i e^{i(-\nu +\theta )}X-i e^{i(\nu
+\theta )}+e^{i(\nu -\theta )}\right]\phi _2^{\left(t_0\right)}\\
\end{aligned}
\end{equation}
where the bar is to indicate that we have encircled the pole at the
origin. Connecting to point $b$:
\begin{equation}\label{eq55}
\begin{aligned}
\psi _{\bar{b}} & =\left[\left(2e^{-i \nu }+e^{i \nu }+e^{-3i \nu
}\right)e^{-i \Gamma }X+e^{i(-\nu +\Gamma )}X+i \left(e^{-i \nu
}+e^{i \nu }\right)e^{-i (\Gamma +2\theta )}Y\right]e^{i \theta
}\phi
_1^{\left(t_0\right)}\\
& + \left[i\left(e^{-i \Gamma } e^{-2i \nu -i \Gamma }+e^{i \Gamma
}\right)e^{i(\theta -\nu )}X-e^{-i(\nu +\Gamma +\theta
)}Y\right]\phi
_2^{\left(t_0\right)}\\
\end{aligned}
\end{equation}
switching to reference point $t_1$:
\begin{equation}\label{eq56}
\begin{aligned}
\psi _{\bar{b}}&=\left[\left(2+e^{2i \nu }+e^{-2i \nu }\right)e^{-i
\Gamma }X+e^{i \Gamma }X+i \left(1+e^{2i \nu }\right)e^{-i (\Gamma
+2\theta )}Y\right]e^{i \theta }\phi
_1^{\left(t_1\right)}\\
& +\left[i\left(e^{-i \Gamma }+e^{-2i \nu -i \Gamma }+e^{i \Gamma
}\right)e^{i(\theta -2\nu )}X-e^{-i(2\nu +\Gamma +\theta
)}Y\right]\phi _2^{\left(t_1\right)}
\end{aligned}
\end{equation}
and finally returning to point $a$:
\begin{equation}\label{eq57}
\begin{aligned}
\psi _{\bar{a}}& =e^{-i \Gamma }\left[e^{i \theta }\left(2(1+\cos
(2\nu ))+e^{2i \Gamma }\right)X+2i e^{i(\nu -\theta )}\cos
(\nu )Y\right]\phi _1^{\left(t_1\right)}\\
& +e^{-i \Gamma }\left[2i e^{i(\theta -\nu )}\cos  \nu \left(1+e^{2i
\Gamma }+2\cos (2\nu )\right)X-e^{-i \theta
}(1+2\cos (2\nu ))Y\right]\phi _2^{\left(t_1\right)}\\
\end{aligned}
\end{equation}
To find $R(\omega)$ we need to properly impose the boundary
conditions on $X$ and $Y$, where $Y$ will be set to unity, then we
identify the coefficient of $\phi_1$ in $\psi_{\bar a}$ as the same
coefficient appearing in $\psi_a$ except that because of the trip
around the contour we have gained a phase:
\begin{equation}\label{eq58}
e^{-i \Gamma }\left[e^{i \theta }\left(2(1+\cos (2\nu ))+e^{2i
\Gamma }\right)R+2i e^{i(\nu -\theta )}\cos (\nu )\right]=e^{i
\theta }X e^{-i \Gamma }
\end{equation}
Solving for $R$ gives us the desired result:
\begin{equation}\label{eq59}
R(\omega )=-e^{i(\nu -2\theta )}\frac{2i \cos  \nu }{e^{\beta \omega
}+1+2\cos (2\nu )}
\end{equation}
To find the corresponding $T$, we notice that the boundary condition
at the horizon, with $\phi_2$ dominating can be connected to point
$d$, which was reached by moving along an anti-Stokes line along
which $\phi_2$ is dominant when $\omega\to -i\infty$. Therefore, we
can make the following identification:
\begin{equation}\label{eq60}
\psi_d = \cdots + T e^{-i \theta} \phi_2^{(t_3)}e^{-i\Gamma}
\end{equation}
comparing with the expression obtained in (\ref{eq53}), we can solve
for $T(\omega)$:
\begin{equation}\label{eq61}
T(\omega )=\frac{e^{\beta  \omega }-1}{1+2\cos (2\nu )+e^{\beta
\omega }}
\end{equation}
For the case of $-\omega$, we just flip the roles of $X$ and $Y$, so
that $Y=R(-\omega)$ is now obtained as follows:
\begin{equation}\label{eq62}
\psi _{\bar{a}}= \cdots +R(-\omega )e^{-i \theta }\phi
_2^{\left(t_1\right)}e^{i \Gamma }
\end{equation}
with $X\equiv 1$. By comparing with (\ref{eq57}), we find:
\begin{equation}\label{eq63}
R(-\omega )=2i e^{i(2\theta -\nu )}\cos  \nu
\end{equation}
Likewise, to find $T(-\omega)$, we make the identification
\begin{equation}\label{eq64}
\psi _d=T e^{-i \theta }\phi _1^{\left(t_3\right)}e^{-i \Gamma
}+\cdots
\end{equation}
and by comparing with the expression in (\ref{eq53}) we find that:
\begin{equation}\label{eq65}
T(-\omega)=1
\end{equation}
The grey-body factor that results from this reads:
\begin{equation}\label{eq66}
\gamma(\omega) =T(\omega )T(-\omega )=\frac{e^{\beta  \omega
}-1}{1+2\cos (2\nu )+e^{\beta  \omega }}
\end{equation}
This result is consistent with the asymptotic quasinormal
frequencies found by Cho \cite{bibCho1} in 4D, and it also implies
Neitzke's results \cite{bibNeitzke} for scalar perturbation around a
Schwarzschild hole for dimensions $D\geq 4$, where the only
modification due to extra dimensions appears through the Hawking
temperature (cf. eq. (\ref{eq2})). Further, in the 6D model that
includes tension, one finds that an additional modification appears
(cf. eq. (\ref{eq5})).

Moving on to brane-localized emissions, we need to use eq.
(\ref{eq36}) as the radial master equation. This would describe
scalar emissions ($s=0$), massless Dirac fermions ($s=\frac{1}{2}$),
and gauge photons ($s=1$). To apply our method, we need to put that
equation in the Schr\"{o}dinger form, in which case the potential
looks like this \cite{bibJungPark}:
\begin{equation}\label{eq67}
\begin{aligned}
  V_s(r)  & = \frac{f}
{{r^2 }}\left[ {\mathcal{A}_{js}  + q_{ns} (1 - f) + \frac{{s^2 }}
{{4f}}\left\{ {(n + 1)(1 - f) - 2f} \right\}^2  + sf} \right] \hfill \\
   & + \frac{{i\omega s}}
{r}\left\{ {(n + 1)(1 - f) - 2f} \right\} \hfill \\
\end{aligned}
\end{equation}
where $$q_{ns}  = (2s + n + 1)(ns + s + 1) - (s/2)(n + 1)(n + 4)$$
and
$$\mathcal{A}_{js}  = j(j + 1) - s(s + 1)$$
To also talk about the graviton localized to the brane, we have to
use another potential (\textit{e.g} \cite{bibPark}):
\begin{equation}\label{eq69}
V_G (r) = \frac{f} {{r^2 }}\left[ {j(j + 1) - \left\{ {(n + 1)^2  +
2} \right\}(1 - f)} \right]
\end{equation}
However, for our purpose, what we care about is the behavior of
these potentials near $r=0$. Keeping only the most singular term in
both cases, we can combine the two potentials into one expression:
\begin{equation}\label{eq70}
V_{{\text{brane}}}  \sim \sigma _n\frac{{r_h^{2(1 + n)} }} {{r^{2(2
+ n)} }}
\end{equation}
with
\begin{equation}\label{eq71}
\sigma _n  = \left\{ {\begin{array}{*{20}l}
   {\tfrac{1}
{4}s^2 (n - 1)^2  - \tfrac{1} {2}sn(n - 1) - (n + 1),} &
{{\text{for}}\,s = 0,\tfrac{1}
{2},1}  \\
   {(n + 1)^2  + 2,} & {{\text{for}}\,s = 2}  \\

 \end{array} } \right.
\end{equation}
To find the greybody factors in this case, we duplicate our steps
for bulk emissions, and we find that the only difference occurs in
the $\nu$, which we can readily find by comparing the expressions
(\ref{eq70}) and (\ref{eq72}):
\begin{equation}\label{eq73}
\nu _{{\text{brane}}}  =  - \pi \sqrt {\frac{{\sigma _n }} {{(n +
2)^2 }} + \frac{1} {4}}
\end{equation}
In this case, the value of the contour integral
$\Gamma_{\mathrm{brane}}$ is found to be:
\begin{equation}\label{eq75}
\Gamma _{{\text{brane}}}  = \mathop{\int\mkern-20.8mu
\circlearrowright}
 {Q\,{\text{d}}r}  =  - i\pi \sqrt {1 - \left( {\frac{{s(n + 1) + 2i\omega r_h }}
{{n + 1}}} \right)^2 }  \approx  - \frac{1} {2}i\pi \beta
\end{equation}
which is asymptotically the same as $\Gamma _{{\text{bulk}}}$.

These results show that the grey-body factors in
all cases are governed by (\ref{eq66}) with differences coming
through the value of $\nu$. Tables \ref{tab:table1} and
\ref{tab:table2} summarize this for bulk and brane-localized
emissions, respectively.
\begin{table}[ht]
\caption{\label{tab:table1}Expressions for $\nu$ for different bulk
perturbations}
\begin{tabular}{l c}
\hline\hline
Perturbation & $\nu_{\mathrm{bulk}}$ \\ [0.5ex] \hline
scalar \& gravi-tensor & 0 \\
gravi-vector & $- \pi$ \\
EM vector & $ - \frac{1} {{n + 2}}\pi $ \\
EM scalar & $ - \frac{{n + 1}} {{n + 2}}\pi $ \\
gravi-scalar & $ \sim  - \frac{\pi } {2}\left[ {2 + 0.674(n + 4)^{ -
0.5445} } \right]$ \\
Dirac fermion & $ - \frac{\pi }{2}$\\ [1ex]
\hline
\end{tabular}
%\label{table:nonlin}
\end{table}

\begin{table}[ht]
\caption{\label{tab:table2}Expressions for $\nu$ for different
brane-localized perturbations}
\begin{tabular}{c c}
\hline\hline Spin & $\nu_{\mathrm{brane}}$ \\ [0.5ex] \hline
0 & $ -
\frac{{\pi n}} {{2(n + 2)}}$ \\
$\frac{1}{2}$ & $ - \frac{\pi } {{2^{3/2} }}\frac{{\sqrt {1 + n^2 }
}} {{n + 2}}$ \\
1 & $ - \pi \frac{{\sqrt {n(n - 2) + 2} }}
{{2(n + 2)}}$ \\
2 & $ - \pi \sqrt {\frac{{(n + 1)^2  + 2}} {{(n + 2)^2 }} + \frac{1}
{4}} $\\ [1ex] \hline
\end{tabular}
%\label{table:nonlin}
\end{table}
Since the asymptotic quasinormal frequencies are the poles
of the grey-body factors, we see that the standard formula still holds
in form,
\begin{equation}\label{eq77}
\frac{\omega } {{T_{\text{H}} }} = 2\pi i\left( {\tilde n +
\tfrac{1} {2}} \right) + \ln \left( {1 + 2\cos 2\nu } \right)
\end{equation}
where $\tilde n$ is the mode number, however, $\nu$ now
depends on the number of extra dimensions (as already
found in \cite{bibPark}).

As a special case, the addition of tension in an $n=2$ spacetime,
will only modify $T_{\text{H}}$, where it can be seen from equations
(\ref{eq5}) and (\ref{eq2}) that the Hawking temperature decreases
with brane tension,
\begin{equation}\label{eq78}
T_{\mathrm{H}}\to b^{1/3} T_{\mathrm{H}}
\end{equation}
Since this occurs in an exponent (eq.~(\ref{eq66})), even a moderate value of
$b$ will lead to $\gamma(\omega)\to 1$ faster with increasing frequency.

\section{Conclusion}
\label{sec5}

With the coming online of the LHC, the possibility of experimentally
testing many of the theories that go beyond the Standard Model is a
realistic one. Detecting black holes in high-energy collisions, if
observed, would, for the first time, offer the opportunity to probe
the realm of quantum gravity helping us resolve
some of the most important puzzles of 20th century physics. This means that we
need to be well-prepared with realistic predictions to interpret the
possible observations. In the study of the creation and evaporation
of mini black-holes, a main prediction of brane-world models, brane
tension has often been neglected. While predictions may not be
radically modified by including it, the fact that we are trying to
observe something we have never observed before makes it imperative
to strive to be as realistic and accurate as possible. Building on earlier work, we made an attempt to
address this issue.

We explored some of the implications of the codimension-2 model on
grey-body factors for various types of bulk and brane emissions from
a black-hole residing on a brane of finite tension. We saw that the
results presented follow primarily from two modifications: {\em (a)}
the enhancement of the horizon radius, which enhanced the emission
spectra with increasing tension, and {\em (b)} the increase of the
angular eigenvalues with tension, which resulted in amplified
absorption probability on the brane. We calculated analytically
grey-body factors in the low frequency regime. We obtained
expressions that had a non-trivial dependence on the brane tension
with observable implications which we discussed. In particular, we
saw that the observable emission rates got enhanced for both bulk
and brane emission, yet we obtained a higher rate on the brane.
It should be stressed that in the case of rotating black-holes our results
are modified due to super-radiance.

We also obtained analytic results in the case of large imaginary frequency.
We derived expressions
for grey-body factor in the bulk and on the brane for a
Schwarzschild black hole in an arbitrary number of dimensions. The special
case of two extra dimensions with tension was seen to follow from a
simple modification of the Hawking temperature.

We compared our analytic expressions in the two asymptotic regimes of low and high frequencies, respectively, with exact numerical solutions and discussed the
range of intermediate frequencies where our analytic expressions are not accurate.
The size of this intermediate range varies depending on the different parameters of the setup.
Interestingly, this range is shifted toward high frequencies with increasing tension indicating a wider range of validity of our analytic low frequency expressions.

In conclusion, our results show that brane tension has a
pronounced effect on predictions. Although the model we discussed is
not the most general one, it offers the opportunity to explore both
qualitatively and quantitatively the effects of adding tension. It would be interesting to go beyond the model considered here
generalizing it to other geometries and an arbitrary number of
extra dimensions. We hope to report on progress in this direction soon.

\section*{Acknowledgment}

We thank Dejan Stojkovic for discussions.
Work supported in part by the Department of Energy under grant DE-FG05-91ER40627.

\newpage

\begin{thebibliography}{48}
\expandafter\ifx\csname
natexlab\endcsname\relax\def\natexlab#1{#1}\fi
\expandafter\ifx\csname bibnamefont\endcsname\relax
  \def\bibnamefont#1{#1}\fi
\expandafter\ifx\csname bibfnamefont\endcsname\relax
  \def\bibfnamefont#1{#1}\fi
\expandafter\ifx\csname citenamefont\endcsname\relax
  \def\citenamefont#1{#1}\fi
\expandafter\ifx\csname url\endcsname\relax
  \def\url#1{\texttt{#1}}\fi
\expandafter\ifx\csname urlprefix\endcsname\relax\def\urlprefix{URL
}\fi \providecommand{\bibinfo}[2]{#2}
\providecommand{\eprint}[2][]{\url{#2}}

\bibitem[{\citenamefont{Arkani-Hamed et~al.}(1998)\citenamefont{Arkani-Hamed,
  Dimopoulos, and Dvali}}]{bibADD1}
\bibinfo{author}{\bibfnamefont{N.}~\bibnamefont{Arkani-Hamed}},
  \bibinfo{author}{\bibfnamefont{S.}~\bibnamefont{Dimopoulos}}
  \bibnamefont{and} \bibinfo{author}{\bibfnamefont{G.~R.} \bibnamefont{Dvali}},
  \bibinfo{journal}{Phys.\ Lett.\ B} \textbf{\bibinfo{volume}{429}},
  \bibinfo{pages}{263} (\bibinfo{year}{1998}),
  \bibinfo{note}{\texttt{arXiv:hep-ph/9803315}}.

\bibitem[{\citenamefont{Antoniadis et~al.}(1998)\citenamefont{Antoniadis,
  Arkani-Hamed, Dimopoulos, and Dvali}}]{bibADD2}
\bibinfo{author}{\bibfnamefont{I.}~\bibnamefont{Antoniadis}},
  \bibinfo{author}{\bibfnamefont{N.}~\bibnamefont{Arkani-Hamed}},
  \bibinfo{author}{\bibfnamefont{S.}~\bibnamefont{Dimopoulos}}
  \bibnamefont{and} \bibinfo{author}{\bibfnamefont{G.~R.} \bibnamefont{Dvali}},
  \bibinfo{journal}{Phys.\ Lett.\ B} \textbf{\bibinfo{volume}{436}},
  \bibinfo{pages}{257} (\bibinfo{year}{1998}),
  \bibinfo{note}{\texttt{arXiv:hep-ph/9804398}}.

\bibitem[{\citenamefont{Arkani-Hamed et~al.}(1999)\citenamefont{Arkani-Hamed,
  Dimopoulos, and Dvali}}]{bibADD3}
\bibinfo{author}{\bibfnamefont{N.}~\bibnamefont{Arkani-Hamed}},
  \bibinfo{author}{\bibfnamefont{S.}~\bibnamefont{Dimopoulos}},
  \bibnamefont{and} \bibinfo{author}{\bibfnamefont{G.~R.} \bibnamefont{Dvali}},
  \bibinfo{journal}{Phys.\ Rev.\ D} \textbf{\bibinfo{volume}{59}},
  \bibinfo{pages}{086004} (\bibinfo{year}{1999}),
  \bibinfo{note}{\texttt{arXiv:hep-ph/9807344}}.

\bibitem[{\citenamefont{Randall and Sundrum}(1999{\natexlab{a}})}]{bibRS1}
\bibinfo{author}{\bibfnamefont{L.}~\bibnamefont{Randall}} \bibnamefont{and}
  \bibinfo{author}{\bibfnamefont{R.}~\bibnamefont{Sundrum}},
  \bibinfo{journal}{Phys.\ Rev.\ Lett.} \textbf{\bibinfo{volume}{83}},
  \bibinfo{pages}{3370} (\bibinfo{year}{1999}{\natexlab{a}}),
  \bibinfo{note}{\texttt{arXiv:hep-ph/9905221}}.

\bibitem[{\citenamefont{Randall and Sundrum}(1999{\natexlab{b}})}]{bibRS2}
\bibinfo{author}{\bibfnamefont{L.}~\bibnamefont{Randall}} \bibnamefont{and}
  \bibinfo{author}{\bibfnamefont{R.}~\bibnamefont{Sundrum}},
  \bibinfo{journal}{Phys.\ Rev.\ Lett.} \textbf{\bibinfo{volume}{83}},
  \bibinfo{pages}{4690} (\bibinfo{year}{1999}{\natexlab{b}}),
  \bibinfo{note}{\texttt{arXiv:hep-ph/9906064}}.

\bibitem[{\citenamefont{Banks and Fischler}(1999)}]{bibBanks}
\bibinfo{author}{\bibfnamefont{T.}~\bibnamefont{Banks}} \bibnamefont{and}
  \bibinfo{author}{\bibfnamefont{W.}~\bibnamefont{Fischler}},
%  (\bibinfo{year}{1999}),
\eprint{\texttt{arXiv:hep-th/9906038}}.

\bibitem[{\citenamefont{Argyres et~al.}(1998)\citenamefont{Argyres, Dimopoulos,
  , and March-Russell}}]{bibADM}
\bibinfo{author}{\bibfnamefont{P.~C.} \bibnamefont{Argyres}},
  \bibinfo{author}{\bibfnamefont{S.}~\bibnamefont{Dimopoulos}}
  \bibnamefont{and}
  \bibinfo{author}{\bibfnamefont{J.}~\bibnamefont{March-Russell}},
  \bibinfo{journal}{Phys.\ Lett.\ B} \textbf{\bibinfo{volume}{441}}, \bibinfo{pages}{96} (\bibinfo{year}{1998}),
  \bibinfo{note}{\texttt{arXiv:hep-th/9808138}}.

\bibitem[{\citenamefont{Dimopoulos and Landsberg}(2001)}]{bibDL}
\bibinfo{author}{\bibfnamefont{S.}~\bibnamefont{Dimopoulos}} \bibnamefont{and}
  \bibinfo{author}{\bibfnamefont{G.}~\bibnamefont{Landsberg}},
  \bibinfo{journal}{Phys.\ Rev.\ Lett.} \textbf{\bibinfo{volume}{87}},
  \bibinfo{pages}{161602} (\bibinfo{year}{2001}),
  \bibinfo{note}{\texttt{arXiv:hep-ph/0106295}}.

\bibitem[{\citenamefont{Giddings and Thomas}(2002)}]{bibGT}
\bibinfo{author}{\bibfnamefont{S.~B.} \bibnamefont{Giddings}} \bibnamefont{and}
  \bibinfo{author}{\bibfnamefont{S.}~\bibnamefont{Thomas}},
  \bibinfo{journal}{Phys.\ Rev.\ D} \textbf{\bibinfo{volume}{65}},
  \bibinfo{pages}{056010} (\bibinfo{year}{2002}),
  \bibinfo{note}{\texttt{arXiv:hep-ph/0106219}}.

\bibitem[{\citenamefont{Kokkotas and Schmidt}(1999)}]{bibKS}
\bibinfo{author}{\bibfnamefont{K.~D.} \bibnamefont{Kokkotas}} \bibnamefont{and}
  \bibinfo{author}{\bibfnamefont{B.}~\bibnamefont{Schmidt}},
  \bibinfo{journal}{Living Rev.\ Rel.} \textbf{\bibinfo{volume}{2}},
  \bibinfo{pages}{2} (\bibinfo{year}{1999}),
  \bibinfo{note}{\texttt{arXiv:gr-qc/9909058}}.

\bibitem[{\citenamefont{Kanti}(2004)}]{bibSM2}
\bibinfo{author}{\bibfnamefont{P.}~\bibnamefont{Kanti}},
  \bibinfo{journal}{Int.\ J.\ Mod.\ Phys.\ A} \textbf{\bibinfo{volume}{19}},
  \bibinfo{pages}{4899} (\bibinfo{year}{2004}),
  \bibinfo{note}{\texttt{arXiv:hep-ph/0402168}}.

\bibitem[{\citenamefont{Nollert}(199)}]{bibNoll}
\bibinfo{author}{\bibfnamefont{H.}~\bibnamefont{Nollert}},
  \bibinfo{journal}{Class.\ Quantum Grav.} \textbf{\bibinfo{volume}{16}},
  \bibinfo{pages}{R159} (\bibinfo{year}{1999}).

\bibitem[{\citenamefont{Nat\'{a}rio and Schiappa}(2004)}]{bibNatario}
\bibinfo{author}{\bibfnamefont{J.}~\bibnamefont{Nat\'{a}rio}} \bibnamefont{and}
  \bibinfo{author}{\bibfnamefont{R.}~\bibnamefont{Schiappa}},
  \bibinfo{journal}{Adv.\ Theor.\ Math.\ Phys.} \textbf{\bibinfo{volume}{8}},
  \bibinfo{pages}{1001} (\bibinfo{year}{2004}),
  \bibinfo{note}{\texttt{arXiv:hep-th/0411267}}.

\bibitem[{\citenamefont{Harmark et~al.}(2007)\citenamefont{Harmark,
  Nat\'{a}rio, and Schiappa}}]{bibHarmark}
\bibinfo{author}{\bibfnamefont{T.}~\bibnamefont{Harmark}},
  \bibinfo{author}{\bibfnamefont{J.}~\bibnamefont{Nat\'{a}rio}},
  \bibnamefont{and} \bibinfo{author}{\bibfnamefont{R.}~\bibnamefont{Schiappa}},
  %(\bibinfo{year}{2007}),
\bibinfo{note}{CERN-PH-TH/2007-091},
  \eprint{\texttt{arXiv:0708.0017}}.

\bibitem[{\citenamefont{Myers and Perry}(1986)}]{bibMyers}
\bibinfo{author}{\bibfnamefont{R.}~\bibnamefont{Myers}} \bibnamefont{and}
  \bibinfo{author}{\bibfnamefont{M.}~\bibnamefont{Perry}},
  \bibinfo{journal}{Annals\ Phys.} \textbf{\bibinfo{volume}{172}},
  \bibinfo{pages}{304} (\bibinfo{year}{1986}).

\bibitem[{\citenamefont{Dimopoulos~al.}(2001)\citenamefont{Dimopoulos, Landsberg}}]{Snowmass}
\bibinfo{author}{\bibfnamefont{S.} \bibnamefont{Dimopoulos}} \bibnamefont{and}
  \bibinfo{author}{\bibfnamefont{G.}~\bibnamefont{Landsberg}}, in
  \emph{
  \bibinfo{title}{Proc. International Workshop on Future of Particle Physics, Snowmass
2001}}, (\bibinfo{year}{2001}),
  \bibinfo{note}{SNOWMASS-2001-P321}.

\bibitem[{\citenamefont{Harris et~al.}(2003)}]{bibHRW}
\bibinfo{author}{\bibfnamefont{C. M.}~\bibnamefont{Harris}},
\bibinfo{author}{\bibfnamefont{P.}~\bibnamefont{Richardson}} \bibnamefont{and}
  \bibinfo{author}{\bibfnamefont{B. R.}~\bibnamefont{Webber}},
  \bibinfo{journal}{JHEP} \textbf{\bibinfo{volume}{0308}}, \bibinfo{pages}{033}
  (\bibinfo{year}{2003}), \bibinfo{note}{\texttt{arXiv:hep-ph/0307305}}.

\bibitem[{\citenamefont{Cavaglia et~al.}(2006)}]{bibCGCS}
\bibinfo{author}{\bibfnamefont{M.}~\bibnamefont{Cavaglia}},
\bibinfo{author}{\bibfnamefont{R.}~\bibnamefont{Godang}},
\bibinfo{author}{\bibfnamefont{L.}~\bibnamefont{Cremaldi}} \bibnamefont{and}
  \bibinfo{author}{\bibfnamefont{D.}~\bibnamefont{Summers}},
  \bibinfo{journal}{Comput.\ Phys.\ Commun.} \textbf{\bibinfo{volume}{177}}, \bibinfo{pages}{506}
  (\bibinfo{year}{2007}), \bibinfo{note}{\texttt{arXiv:hep-ph/0609001}}.

\bibitem[{\citenamefont{Dai et~al.}(2008)}]{bibDSSC}
\bibinfo{author}{\bibfnamefont{D.~C.}~\bibnamefont{Dai}},
\bibinfo{author}{\bibfnamefont{G.~D.}~\bibnamefont{Starkman}},
\bibinfo{author}{\bibfnamefont{D.}~\bibnamefont{Stojkovic}} \bibnamefont{and}
  \bibinfo{author}{\bibfnamefont{C.}~\bibnamefont{Issever}},
  \bibinfo{journal}{Phys.\ Rev.\ D} \textbf{\bibinfo{volume}{77}}, \bibinfo{pages}{076007}
  (\bibinfo{year}{2008}), \bibinfo{note}{\texttt{arXiv:0711.3012}}.

\bibitem[{\citenamefont{Kaloper and Kiley}(2006)}]{bibKK}
\bibinfo{author}{\bibfnamefont{N.}~\bibnamefont{Kaloper}} \bibnamefont{and}
  \bibinfo{author}{\bibfnamefont{D.}~\bibnamefont{Kiley}},
  \bibinfo{journal}{JHEP} \textbf{\bibinfo{volume}{0603}}, \bibinfo{pages}{77}
  (\bibinfo{year}{2006}), \bibinfo{note}{\texttt{arXiv:hep-th/0601110}}.

\bibitem[{\citenamefont{Kiley}(2007)}]{bibKiley}
\bibinfo{author}{\bibfnamefont{D.}~\bibnamefont{Kiley}},
  \bibinfo{journal}{Phys.\ Rev.\ D} \textbf{\bibinfo{volume}{76}},
  \bibinfo{pages}{126002} (\bibinfo{year}{2007}),
  \bibinfo{note}{\texttt{arXiv:0708.1016}}.

\bibitem[{\citenamefont{Dai et~al.}(2007)\citenamefont{Dai, Kaloper, Starkman,
  and Stojkovic}}]{bibDKSS}
\bibinfo{author}{\bibfnamefont{D.~C.} \bibnamefont{Dai}},
  \bibinfo{author}{\bibfnamefont{N.}~\bibnamefont{Kaloper}},
  \bibinfo{author}{\bibfnamefont{G.~D.}~\bibnamefont{Starkman}} \bibnamefont{and}
  \bibinfo{author}{\bibfnamefont{D.}~\bibnamefont{Stojkovic}},
  \bibinfo{journal}{Phys.\ Rev.\ D} \textbf{\bibinfo{volume}{75}},
  \bibinfo{pages}{024043} (\bibinfo{year}{2007}),
  \bibinfo{note}{\texttt{arXiv:hep-th/0611184}}.

\bibitem[{\citenamefont{Chen et~al.}(2007)\citenamefont{Chen, Wang, and
  Su}}]{bibCWS}
\bibinfo{author}{\bibfnamefont{S.}~\bibnamefont{Chen}},
  \bibinfo{author}{\bibfnamefont{B.}~\bibnamefont{Wang}} \bibnamefont{and}
  \bibinfo{author}{\bibfnamefont{R.-K.} \bibnamefont{Su}},
  \bibinfo{journal}{Phys.\ Lett.\ B} \textbf{\bibinfo{volume}{647}},
  \bibinfo{pages}{282} (\bibinfo{year}{2007}),
  \bibinfo{note}{\texttt{arXiv:hep-th/0701209}}.

\bibitem[{\citenamefont{Chen et~al.}(2008)\citenamefont{Chen, Wang, Su, , and
  Hwang}}]{bibCWSWP}
\bibinfo{author}{\bibfnamefont{S.}~\bibnamefont{Chen}},
  \bibinfo{author}{\bibfnamefont{B.}~\bibnamefont{Wang}},
  \bibinfo{author}{\bibfnamefont{R.}~\bibnamefont{Su}} \bibnamefont{and}
  \bibinfo{author}{\bibfnamefont{W.-Y.~P.} \bibnamefont{Hwang}},
  \bibinfo{journal}{JHEP} \textbf{\bibinfo{volume}{0803}}, \bibinfo{pages}{019}
  (\bibinfo{year}{2008}),
  \bibinfo{note}{\texttt{arXiv:0711.3599}}.

\bibitem[{\citenamefont{Kobayashi et~al.}(2008)\citenamefont{Kobayashi, Nozawa, and Takamizu}}]{bibKNT}
\bibinfo{author}{\bibfnamefont{T.} \bibnamefont{Kobayashi}},
  \bibinfo{author}{\bibfnamefont{M.}~\bibnamefont{Nozawa}} \bibnamefont{and}
  \bibinfo{author}{\bibfnamefont{Y.}~\bibnamefont{Takamizu}},
  \bibinfo{journal}{Phys.\ Rev.\ D} \textbf{\bibinfo{volume}{77}},
  \bibinfo{pages}{044022} (\bibinfo{year}{2008}),
  \bibinfo{note}{\texttt{arXiv:0711.1395}}.

\bibitem[{\citenamefont{Cho et~al.}(2007)\citenamefont{Cho, Cornell, Doukas, ,
  and Naylor}}]{bibCCDN4}
\bibinfo{author}{\bibfnamefont{H.~T.} \bibnamefont{Cho}},
  \bibinfo{author}{\bibfnamefont{A.~S.} \bibnamefont{Cornell}},
  \bibinfo{author}{\bibfnamefont{J.}~\bibnamefont{Doukas}} \bibnamefont{and}
  \bibinfo{author}{\bibfnamefont{W.}~\bibnamefont{Naylor}}, in
  \emph{
  \bibinfo{title}{Proceedings of 17th Workshop on General Relativity and Gravitation in
  Japan}}, (\bibinfo{year}{2008}),
  \bibinfo{note}{\texttt{arXiv:0803.2547}}.

\bibitem[{\citenamefont{P.~Kanti}(2002)}]{bibKantiI}
\bibinfo{author}{\bibnamefont{P.~Kanti} \bibfnamefont{and J.~March-Russell}},
  \bibinfo{journal}{Phys.\ Rev.\ D} \textbf{\bibinfo{volume}{66}},
  \bibinfo{pages}{024023} (\bibinfo{year}{2002}),
  \bibinfo{note}{\texttt{arXiv:hep-ph/0203223}}.

\bibitem[{\citenamefont{P.~Kanti}(2003)}]{bibKantiII}
\bibinfo{author}{\bibnamefont{P.~Kanti} \bibfnamefont{and J.~March-Russell}},
  \bibinfo{journal}{Phys.\ Rev.\ D} \textbf{\bibinfo{volume}{67}},
  \bibinfo{pages}{104019} (\bibinfo{year}{2003}),
  \bibinfo{note}{\texttt{arXiv:hep-ph/0212199}}.

\bibitem[{\citenamefont{al~Binni and Siopsis}(2007)}]{bibPaper1}
\bibinfo{author}{\bibfnamefont{U.~A.} \bibnamefont{al~Binni}} \bibnamefont{and}
  \bibinfo{author}{\bibfnamefont{G.}~\bibnamefont{Siopsis}},
  \bibinfo{journal}{Phys.\ Rev.\ D} \textbf{\bibinfo{volume}{76}},
  \bibinfo{pages}{104031} (\bibinfo{year}{2007}),
  \bibinfo{note}{\texttt{arXiv:0708.3363}}.

\bibitem[{\citenamefont{Cho et~al.}(2008{\natexlab{a}})\citenamefont{Cho,
  Cornell, Doukas, , and Naylor}}]{bibCCDN1}
\bibinfo{author}{\bibfnamefont{H.~T.} \bibnamefont{Cho}},
  \bibinfo{author}{\bibfnamefont{A.~S.} \bibnamefont{Cornell}},
  \bibinfo{author}{\bibfnamefont{J.}~\bibnamefont{Doukas}} \bibnamefont{and}
  \bibinfo{author}{\bibfnamefont{W.}~\bibnamefont{Naylor}},
  \bibinfo{journal}{Phys.\ Rev.\ D} \textbf{\bibinfo{volume}{77}},
  \bibinfo{pages}{041502} (\bibinfo{year}{2008}),
  \bibinfo{note}{\texttt{arXiv:0710.5267}}.

\bibitem[{\citenamefont{Creek et~al.}(2006)\citenamefont{Creek, Efthimiou,
  Kanti, and Tamvakis}}]{bibCEKT}
\bibinfo{author}{\bibfnamefont{S.}~\bibnamefont{Creek}},
  \bibinfo{author}{\bibfnamefont{O.}~\bibnamefont{Efthimiou}},
  \bibinfo{author}{\bibfnamefont{P.}~\bibnamefont{Kanti}} \bibnamefont{and}
  \bibinfo{author}{\bibfnamefont{K.}~\bibnamefont{Tamvakis}},
  \bibinfo{journal}{Phys. Lett. B} \textbf{\bibinfo{volume}{635}},
  \bibinfo{pages}{39} (\bibinfo{year}{2006}),
  \bibinfo{note}{\texttt{arXiv:hep-th/0601126}}.

\bibitem[{\citenamefont{Cardoso
  et~al.}(2006{\natexlab{a}})\citenamefont{Cardoso, Cavaglia, and
  Gualtieri}}]{bibCCG1}
\bibinfo{author}{\bibfnamefont{V.}~\bibnamefont{Cardoso}},
  \bibinfo{author}{\bibfnamefont{M.}~\bibnamefont{Cavaglia}} \bibnamefont{and}
  \bibinfo{author}{\bibfnamefont{L.}~\bibnamefont{Gualtieri}},
  \bibinfo{journal}{JHEP} \textbf{\bibinfo{volume}{0602}}, \bibinfo{pages}{021}
  (\bibinfo{year}{2006}{\natexlab{a}}),
  \bibinfo{note}{\texttt{arXiv:hep-th/0512116}}.

\bibitem[{\citenamefont{Ida et~al.}(2003)\citenamefont{Ida, Oda, and
  Park}}]{bibIda1}
\bibinfo{author}{\bibfnamefont{D.}~\bibnamefont{Ida}},
  \bibinfo{author}{\bibfnamefont{K.~Y.}~\bibnamefont{Oda}} \bibnamefont{and}
  \bibinfo{author}{\bibfnamefont{S.~C.} \bibnamefont{Park}},
  \bibinfo{journal}{Phys.\ Rev.\ D} \textbf{\bibinfo{volume}{67}},
  \bibinfo{pages}{064025} (\bibinfo{year}{2003}),
  \bibinfo{note}{erratum-ibid. \textbf{69}, 049901 (2004)},
\eprint{\texttt{arXiv:hep-th/0212108}}.

\bibitem[{\citenamefont{Ida et~al.}(2005)\citenamefont{Ida, Oda, and
  Park}}]{bibIda2}
\bibinfo{author}{\bibfnamefont{D.}~\bibnamefont{Ida}},
  \bibinfo{author}{\bibfnamefont{K.~Y.}~\bibnamefont{Oda}} \bibnamefont{and}
  \bibinfo{author}{\bibfnamefont{S.~C.} \bibnamefont{Park}},
  %(\bibinfo{year}{2005}),
\eprint{\texttt{arXiv:hep-ph/0501210}}.

\bibitem[{\citenamefont{Ishibashi and Kodama}(2003)}]{bibIK}
\bibinfo{author}{\bibfnamefont{A.}~\bibnamefont{Ishibashi}} \bibnamefont{and}
  \bibinfo{author}{\bibfnamefont{H.}~\bibnamefont{Kodama}},
  \bibinfo{journal}{Prog.\ Theor.\ Phys.} \textbf{\bibinfo{volume}{110}},
  \bibinfo{pages}{701} (\bibinfo{year}{2003}),
\eprint{\texttt{arXiv:hep-th/0305147}}.

\bibitem[{\citenamefont{Cardoso
  et~al.}(2006{\natexlab{b}})\citenamefont{Cardoso, Cavaglia, and
  Gualtieri}}]{bibCCG2}
\bibinfo{author}{\bibfnamefont{V.}~\bibnamefont{Cardoso}},
  \bibinfo{author}{\bibfnamefont{M.}~\bibnamefont{Cavaglia}} \bibnamefont{and}
  \bibinfo{author}{\bibfnamefont{L.}~\bibnamefont{Gualtieri}},
  \bibinfo{journal}{Phys.\ Rev.\ Lett.} \textbf{\bibinfo{volume}{96}},
  \bibinfo{pages}{071301} (\bibinfo{year}{2006}{\natexlab{b}}),
  \bibinfo{note}{erratum-ibid. \textbf{96}, 219902 (2006)},
\eprint{\texttt{arXiv:hep-th/0512002}}.

\bibitem[{\citenamefont{Cvetic and Larsen}(1998)}]{bibCL}
\bibinfo{author}{\bibfnamefont{M.}~\bibnamefont{Cvetic}} \bibnamefont{and}
  \bibinfo{author}{\bibfnamefont{F.}~\bibnamefont{Larsen}},
  \bibinfo{journal}{Phys.\ Rev.\ D} \textbf{\bibinfo{volume}{57}},
  \bibinfo{pages}{6297} (\bibinfo{year}{1998}),
\eprint{\texttt{arXiv:hep-th/9712118}}.

\bibitem[{\citenamefont{Jung and Park}(2007)}]{bibJungPark}
\bibinfo{author}{\bibfnamefont{E.}~\bibnamefont{Jung}} \bibnamefont{and}
  \bibinfo{author}{\bibfnamefont{D.~K.} \bibnamefont{Park}},
  \bibinfo{journal}{Nucl.\ Phys.\ B} \textbf{\bibinfo{volume}{766}},
  \bibinfo{pages}{269} (\bibinfo{year}{2007}),
  \bibinfo{note}{\texttt{arXiv:hep-th/0610089}}.

\bibitem[{\citenamefont{Emparan et~al.}(2000)\citenamefont{Emparan, Horowitz,
  and Myers}}]{bibbh3}
\bibinfo{author}{\bibfnamefont{R.}~\bibnamefont{Emparan}},
  \bibinfo{author}{\bibfnamefont{G.~T.} \bibnamefont{Horowitz}}
  \bibnamefont{and} \bibinfo{author}{\bibfnamefont{R.~C.} \bibnamefont{Myers}},
  \bibinfo{journal}{Phys.\ Rev.\ Lett.} \textbf{\bibinfo{volume}{85}},
  \bibinfo{pages}{499} (\bibinfo{year}{2000}),
  \bibinfo{note}{\texttt{arXiv:hep-th/0003118}}.

\bibitem[{\citenamefont{Frolov and Stojkovic}(2002)}]{bibFS1}
\bibinfo{author}{\bibfnamefont{V. P.}~\bibnamefont{Frolov}} \bibnamefont{and}
  \bibinfo{author}{\bibfnamefont{D.}~\bibnamefont{Stojkovic}},
  \bibinfo{journal}{Phys.\ Rev.\ Lett.} \textbf{\bibinfo{volume}{89}},
  \bibinfo{pages}{151302} (\bibinfo{year}{2002}),
  \bibinfo{note}{\texttt{arXiv:hep-th/0208102}}.

\bibitem[{\citenamefont{Frolov and Stojkovic}(2002)}]{bibFS2}
\bibinfo{author}{\bibfnamefont{V. P.}~\bibnamefont{Frolov}} \bibnamefont{and}
  \bibinfo{author}{\bibfnamefont{D.}~\bibnamefont{Stojkovic}},
  \bibinfo{journal}{Phys.\ Rev.\ D} \textbf{\bibinfo{volume}{66}},
  \bibinfo{pages}{084002} (\bibinfo{year}{2002}),
  \bibinfo{note}{\texttt{arXiv:hep-th/0206046}}.

\bibitem[{\citenamefont{Stojkovic}(2002)}]{bibSt}
  \bibinfo{author}{\bibfnamefont{D.}~\bibnamefont{Stojkovic}},
  \bibinfo{journal}{Phys.\ Rev.\ Lett.} \textbf{\bibinfo{volume}{94}},
  \bibinfo{pages}{011603} (\bibinfo{year}{2005}),
  \bibinfo{note}{\texttt{arXiv:hep-ph/0409124}}.

\bibitem[{\citenamefont{Andersson and Howls}(2004)}]{bibAndHowls}
\bibinfo{author}{\bibfnamefont{N.}~\bibnamefont{Andersson}} \bibnamefont{and}
  \bibinfo{author}{\bibfnamefont{C.}~\bibnamefont{Howls}},
  \bibinfo{journal}{Class.\ Quant.\ Grav.} \textbf{\bibinfo{volume}{21}},
  \bibinfo{pages}{1623} (\bibinfo{year}{2004}),
\eprint{\texttt{arXiv:gr-qc/0307020}}.

\bibitem[{\citenamefont{Cho}(2006)}]{bibCho1}
\bibinfo{author}{\bibfnamefont{H.~T.} \bibnamefont{Cho}},
  \bibinfo{journal}{Phys.\ Rev.\ D} \textbf{\bibinfo{volume}{73}},
  \bibinfo{pages}{024019} (\bibinfo{year}{2006}),
  \bibinfo{note}{\texttt{arXiv:gr-qc/0512052}}.

\bibitem[{\citenamefont{Motl and Neitzke}(2003)}]{bibMotlN}
\bibinfo{author}{\bibfnamefont{L.}~\bibnamefont{Motl}} \bibnamefont{and}
  \bibinfo{author}{\bibfnamefont{A.}~\bibnamefont{Neitzke}},
  \bibinfo{journal}{Adv.\ Theor.\ Math.\ Phys.} \textbf{\bibinfo{volume}{7}},
  \bibinfo{pages}{307} (\bibinfo{year}{2003}),
  \bibinfo{note}{\texttt{arXiv:hep-th/0301173}}.

\bibitem[{\citenamefont{Musiri and Siopsis}(2003)}]{bibMuSi}
\bibinfo{author}{\bibfnamefont{S.}~\bibnamefont{Musiri}} \bibnamefont{and}
  \bibinfo{author}{\bibfnamefont{G.}~\bibnamefont{Siopsis}},
  \bibinfo{journal}{Class.\ Quantum Grav.} \textbf{\bibinfo{volume}{20}},
  \bibinfo{pages}{L285} (\bibinfo{year}{2003}),
  \bibinfo{note}{\texttt{arXiv:hep-th/0308168}}.

\bibitem[{\citenamefont{van~den Brink}(2004)}]{bibAlec}
\bibinfo{author}{\bibfnamefont{A.~M.} \bibnamefont{van~den Brink}},
  \bibinfo{journal}{J.\ Math.\ Phys.} \textbf{\bibinfo{volume}{45}},
  \bibinfo{pages}{327} (\bibinfo{year}{2004}),
  \bibinfo{note}{\texttt{arXiv:gr-qc/0304092}}.

\bibitem[{\citenamefont{Cho et~al.}(2008{\natexlab{b}})\citenamefont{Cho,
  Cornell, Doukas, , and Naylor}}]{bibCCDN2}
\bibinfo{author}{\bibfnamefont{H.~T.} \bibnamefont{Cho}},
  \bibinfo{author}{\bibfnamefont{A.~S.} \bibnamefont{Cornell}},
  \bibinfo{author}{\bibfnamefont{J.}~\bibnamefont{Doukas}} \bibnamefont{and}
  \bibinfo{author}{\bibfnamefont{W.}~\bibnamefont{Naylor}},
  \bibinfo{journal}{Phys.\ Rev.\ D} \textbf{\bibinfo{volume}{77}},
  \bibinfo{pages}{016004} (\bibinfo{year}{2008}{\natexlab{b}}),
  \bibinfo{note}{\texttt{arXiv:0709.1661}}.

\bibitem[{\citenamefont{Cho et~al.}(2007)\citenamefont{Cho, Cornell, Doukas, ,
  and Naylor}}]{bibCCDN3}
\bibinfo{author}{\bibfnamefont{H.~T.} \bibnamefont{Cho}},
  \bibinfo{author}{\bibfnamefont{A.~S.} \bibnamefont{Cornell}},
  \bibinfo{author}{\bibfnamefont{J.}~\bibnamefont{Doukas}} \bibnamefont{and}
  \bibinfo{author}{\bibfnamefont{W.}~\bibnamefont{Naylor}},
  \bibinfo{journal}{Phys.\ Rev.\ D} \textbf{\bibinfo{volume}{75}},
  \bibinfo{pages}{104005} (\bibinfo{year}{2007}),
  \bibinfo{note}{\texttt{arXiv:hep-th/0701193}}.

\bibitem[{\citenamefont{Neitzke}(2003)}]{bibNeitzke}
\bibinfo{author}{\bibfnamefont{A.}~\bibnamefont{Neitzke}},
%  (\bibinfo{year}{2003})
\eprint{\texttt{arXiv:hep-th/0304080}}.

\bibitem[{\citenamefont{Park}(2006)}]{bibPark}
\bibinfo{author}{\bibfnamefont{D.~K.} \bibnamefont{Park}},
  \bibinfo{journal}{Phys.\ Lett.\ B} \textbf{\bibinfo{volume}{633}},
  \bibinfo{pages}{613} (\bibinfo{year}{2006}),
  \bibinfo{note}{\texttt{arXiv:hep-th/0511159}}.

\end{thebibliography}
\end{document}